\documentclass{emulateapj}
\usepackage{latexsym}  % for emulate style
\usepackage{graphicx}
\usepackage{natbib}
\usepackage{color}
\usepackage{amsmath}
\usepackage{hyperref} 
\def\pomega{\varpi}
\def\avrg#1{{\langle #1 \rangle}}
\def\half{{\textstyle{1\over2}}}

\def\Trace{{\rm Tr}} 
\def\th{{\cal T}}
\def\data{{\cal D}}

\def\half{{\textstyle{1\over2}}}
\def\spose#1{\hbox to 0pt{#1\hss}}
\def\lta{\mathrel{\spose{\lower 3pt\hbox{$\mathchar"218$}}
     \raise 2.0pt\hbox{$\mathchar"13C$}}}
\def\gta{\mathrel{\spose{\lower 3pt\hbox{$\mathchar"218$}}
     \raise 2.0pt\hbox{$\mathchar"13E$}}}

\slugcomment{submitted to ApJ}
\shorttitle{Gravitational Wave Detectability with the CMB}
\shortauthors{Farhang et al.}

\begin{document}
\title{Primordial Gravitational Wave Detectability with Deep Small-Sky CMB Experiments}
\author{M. Farhang\altaffilmark{1,2}, J.R. Bond\altaffilmark{1},
  O. Dor{\'e} \altaffilmark{2,3,4}, C.B. Netterfield\altaffilmark{1,5} }
\altaffiltext{1}{ Department of Astronomy and Astrophysics, University of Toronto, 50 St George , Toronto ON, M5S 3H4.}
\altaffiltext{2}{Canadian Institute for Theoretical Astrophysics, 60 St George , Toronto ON, M5S 3H8}
\altaffiltext{3}{Jet Propulsion Laboratory, California Institute of Technology, Pasadena, CA 91109}
\altaffiltext{4}{California Institute of Technology, Pasadena, CA 91125}
 \altaffiltext{5}{Department of Physics, University of Toronto, 60 St. George Street, Toronto, ON M5S 1A7, Canada}
%------------------------------------------
\begin{abstract}
We use Bayesian estimation on direct $T$-$Q$-$U$ CMB polarization maps to  forecast errors on the tensor-to-scalar power ratio $r$, and hence on primordial
gravitational waves,  as a function of sky coverage $f_{\rm sky}$. This
$T$-$Q$-$U$ matrix likelihood filters the quadratic pixel-pixel space into
the optimal combinations needed for $r$ detection for cut skies, providing
enhanced  information over a first-step linear separation into a
combination of $E$, $B$ and mixed modes, and ignoring the latter. With
current computational power and for typical resolutions appropriate
for $r$ detection, the large matrix inversions required are accurate
and fast. Our simulations explore two classes of experiments, with
differing bolometric detector numbers,  sensitivities and
observational strategies. One is motivated by a long duration balloon
experiment like Spider,  with pixel noise $\propto \sqrt{f_{\rm{sky}}}$ for a
specified observing period. This analysis also applies to ground-based
array experiments.  We find that, in the absence of systematic effects
and foregrounds, an experiment with Spider-like noise concentrating on
$f_{\rm{sky}} \sim 0.02$--$0.2$  could place a $2\sigma_r\approx
0.014$ bound ($\sim 95\%$ CL), which rises to $ 0.02$ with an
$\ell$-dependent  foreground residual left over from an assumed
efficient  component separation.  We contrast this with a Planck-like
fixed instrumental noise as $f_{\rm{sky}}$ varies,  which gives
a Galaxy-masked ($f_{\rm sky}=0.75$) $2\sigma_r\approx 0.015$, rising to  $\approx 0.05$ with the
foreground residuals. Using for a figure of merit the  (marginalized)
1D Shannon entropy  of  $r$,  taken relative to  the first  2003 WMAP1
CMB-only constraint, gives $-1.7$ bits from the 2010  WMAP7+ACT data,
$-1.9$ bits from the 2011 WMAP7+SPT data, and forecasts of  -6 bits
from Spider (plus Planck);   this compares with  up to -11 bits for a CMBPol, COrE and PIXIE post-Planck satellites and -13 bits for a perfectly noiseless  cosmic variance limited experiment. We thus confirm the wisdom of the current strategy for $r$ detection of deeply probed patches covering the $f_{\rm{sky}}$ minimum-error trough with balloon and ground experiments.  
\end{abstract}
%-------------------------------------------

\section{Introduction}\label{sec:intro}

Inflation, a period of accelerated expansion in the very early universe, is the most widely accepted scenario to solve the problems of
the otherwise successful standard model of cosmology. In the simplest
models the expansion is driven by an effective potential energy of 
a single scalar field degree of freedom, the inflaton. An unavoidable consequence is the quantum generation of scalar and tensor zero-point fluctuations in
the space-time metric. The former are  3-curvature  perturbations, with associated density fluctuations that can grow via gravitational instability to create the cosmic web, with its rich observational characterization. The latter are gravity waves that induce potentially observable signatures in the spatial structure of the Cosmic Microwave Background (CMB), in particular in its polarization, the focus of this paper. Whereas curl-free $E$-modes of polarization can be produced both by tensor and scalar perturbations,  divergence-free modes of CMB polarization ($B$-modes) would be induced on large scales by primordial 
gravitational waves but not by  scalar curvature fluctuations. Many experiments are in quest of this inflation signature, but the predicted signal, if detectable, is very small and subject to contamination by leakages from the total anisotropy $T$ and from the dominant $E$ polarization, as well as by other systematic effects, so extraordinary care is needed to analyze such data. At  smaller scales, $B$ modes are induced from primordial $E$ modes through gravitational lensing distortions of the CMB polarization patterns,  adding to the complexity of making a clean separation of the tensor-induced signal.  

The primordial scalar and tensor power spectra (fluctuation variances
per $\ln k$) and their ratio $r(k)$ are often approximated by power
laws in the 3D comoving wavenumber $k$, 
\begin{eqnarray}
&& {\cal P}_{\rm s}(k) \approx A_{\rm
    s}(k_{\rm sp})\left(k/k_{\rm sp}\right)^{n_{\rm s}(k_{\rm sp})-1}, \cr
&& {\cal P}_{\rm t}(k) \approx A_{\rm t}(k_{\rm tp})\left(k/k_{\rm
    tp}\right)^{n_{\rm t}(k_{\rm tp})}. \cr
&& r(k)\equiv  {\cal P}_{\rm t}(k) /{\cal P}_{\rm s}(k) \approx r\,
  \left(k/k_{\rm tp}\right)^{n_{\rm t}(k_{\rm tp})-n_{\rm s}(k_{\rm tp})+1} , \nonumber \cr
&& r \equiv r(k_{\rm tp}) \equiv {\cal P}_{\rm t}(k_{\rm tp}) /{\cal
    P}_{\rm s}(k_{\rm tp}) \, .
\nonumber
\end{eqnarray} 
The scalar and tensor pivots $k_{\rm sp}$ and $k_{\rm tp}$ about which the expansions occur are
usually chosen to be different for scalars and tensors to reflect
where the optimal signal weights come from. The main target of many of
the current and coming CMB polarization experiments is, firstly,  a
one-parameter uniform $r$.  An advantage of this ratio  over ${\cal
  P}_t(k_{\rm tp})$ is that it removes a dominant near-degeneracy with the
Thompson depth to Compton scattering $\tau$. The spectrum $r(k)$ also
measures the inflation acceleration history  $\epsilon (a)$, and can
be directly related to the inflaton potential energy through this
relation: 
\begin{eqnarray}
&& r(k)  \approx 16\epsilon(a \approx k/H), \quad \epsilon \equiv -{\rm d}\ln H /{\rm d}\ln a \, , \label{eq:reps} \cr
&& V \approx r M_{\rm P}^4  {\cal P}_{\rm s} 3/2(1-r/48) \sim (10^{16} {\rm Gev})^4 r/0.1\, .
\end{eqnarray} 
Here $M_{\rm P}=1/\sqrt{8\pi G}$ is the reduced Planck mass, with $c$
and $\hbar$ set to unity. The relation $k\approx Ha$, of resolution $k^{-1}$ to the dynamics encoded in the expansion and Hubble
parameters,  $a$ and  $H$, is only approximate of course, but very
useful, e.g., \cite{bon96}.  A detection of $r\sim 0.03-0.2$ would
provide a strong pointer to the specific inflation model.  A tight
upper bound, $r \lesssim 0.03$, would rule out a very large class of
inflation scenarios, a bound that is achievable with the experiments
we explore here. In this paper, we often use $r_{\rm fid}=0.12$ as a
fiducial high-$r$ case for tests, since it is near the 0.13 coming
from the simplest $V=m^2\phi^2/2$ chaotic inflation model, and has an
inflation energy scale $V^{1/4}$ near $10^{16}$ Gev. We also explore
the very small  $r_{\rm fid}<0.01$ regime. 

We would like to learn as much as we can about the full $r(k)$, hence $\epsilon (a)$,  from CMB data.  In addition to the  deviations of the slopes from scale invariance ($n_{\rm t}=0$ and $n_{\rm s}-1=0$), the slopes are expected to "run with $k$" just as the power does, although they may be approximately constant over the observable CMB range. The first order variations in $\ln k$ define  scalar and tensor "running " parameters, the first terms in polynomial  expansions in higher order "running of running" parameters. In this paper $n_{\rm s}(k)$ is not our target, nor are high multipole CMB experiments which are necessary to get the long baseline needed to show whether $n_{\rm s}$ runs or not. 

A consequence of  the fall-off of the tensor-induced  CMB signal
beyond $\ell \sim 150$ is that only limited information can be
obtained on $n_{\rm t}(k)$ --- enough to allow a number of broad bands for
$r(k)$, but  not enough for $n_{\rm t}(k_{\rm tp})$, let alone $n_{\rm t}(k)$,  to be
determined with sufficient accuracy to test well the inflation
consistency relation for gravity waves.   In the limited 2-parameter
tensor parameter space of $r$ and uniform $n_{\rm t}$,  this consistency
condition is  (e.g., \cite{bon96})
\begin{eqnarray}
&& n_{\rm t} \approx - r/8 /(1-r/16) , 
\end{eqnarray}\label{eq:rnt}
so a convincing test would require an order of magnitude better determination of $n_{\rm t}$ than $r$. Another complication in relating the experiments to inflation theory is that there is still observational room for subdominant scalar isocurvature perturbations in addition to the dominant curvature ones when multiple fields are dynamically important during or immediately after inflation; such fields are widely invoked for catalyzing the production of entropy at the end of inflation.  Isocurvature perturbations with a nearly scale invariant primordial spectrum have significantly enhanced low-$\ell$ CMB power because of the isocurvature effect \cite{bon96}, and that region, overlapping with the gravity wave induced CMB power, is where the constraint on the overall isocurvature amplitude comes from \citep{sie07}.

All CMB polarization experiments are
limited in sky coverage by instrumental or Galactic foreground constraints. Thus, even though the $B$ modes provide a unique $r$-signature and are orthogonal to the $E$ modes over the full sky,  realistically mode-mixing must always be dealt with, even though it may be larger for smaller $f_{\rm{sky}}$.   Assessing the  trade offs between shallow large-sky and  deep small-sky observational strategies is the target of our investigation. Going for deep and small has the advantage that one can select the most foreground-free patches to target to decrease the high level of foreground subtraction. As well, the long waves which dominate foregrounds are naturally filtered. Ground-based or balloon-borne experiments using the deep and small-sky strategy are: BICEP and
BICEP2\footnote{http://bicep.caltech.edu/public/} , QUIET\footnote{http://quiet.uchicago.edu/}, PolarBear\footnote{http://bolo.berkeley.edu/polarbear/}, EBEX\footnote{http://groups.physics.umn.edu/cosmology/ebex/}, 
Spider\footnote{http://www.astro.caltech.edu/~lgg/spider/spider\_front.htm},
KECK\citep{she11},
ABS\footnote{http://www.princeton.edu/physics/research/cosmology-experiment/abs-experiment/},
PIPER \citep{chu10}.   Planck (and WMAP) are  (relatively) shallow and large-sky. Proposed next-generation satellite experiments such as COrE \citep{core11}, PIXIE \citep{kog11}  and LiteBIRD \footnote{http://cmbpol.kek.jp/litebird/index.html} are deep and large-sky. 

In this paper, we first review the general Bayesian framework for determining parameters to introduce the notations we use. We cast the quest for $r$ into an information-theoretic language in which the forecasted outcomes of different experiments can be contrasted by considering the differences in their reduced {\it a posteriori} Shannon entropies for $r$, $S_{1{\rm f}}(r\vert {\rm expt})$. We discuss the two basic approaches for constraining cosmological observables,  such as those associated with inflation, and the relation of these to $E$-$B$ mixing: (1) the 
$\ell$-space approach in which CMB maps are first compressed onto power spectrum parameters for $TT$-$TE$-$EE$ and $BB$, which are then compressed onto cosmic parameters;  and (2) direct parameter  extraction of $r$ from map likelihoods. Our primary target is $r$ and not the $B$-mode spectrum, hence the optimal one-step estimation from maps is preferred, provided it is computationally feasible --  which it is for Spider-like experiments. The leakage between the $E$ and $B$ modes and its impact on  $r$ is
quantified in \S~\ref{sec:LRT}.  In \S~\ref{sec:results} we present details of  
the method we use to bypass explicit $E$-$B$ de-mixing and apply it to simulated data for realistic instrumental and foreground-residual noise levels for Spider-like and Planck-like experiments as $f_{\rm{sky}}$ varies. We end with our conclusions from this study. 
%-------------------------------------------

\section{Bayesian CMB Analysis of Maps, Bandpowers and Cosmic Parameters}\label{sec:likelihood}

As has become conventional in CMB analysis, the framework envisaged to
reduce the information from Spider-like raw time ordered data to
constraints on cosmic parameters, in particular our target $r$,  is
one of a long Bayesian chain of conditional probabilities
\citep{bon96,bon01}.  To introduce our notation, we review that framework with polarization. We  also remark on how the associated conditional Shannon entropies decrease as we flow along the Bayesian chain, a novel way of looking at what is being done as the data is reduced to a precious set of parameter bits. 
%-------------------------------------------

\subsection{ Reducing Noisy Data with Bayesian Chains}\label{sec:bayes}
%-------------------------------------------

\subsubsection{The Information Action in Bayesian Chains}\label{sec:infoaction}

  In Bayesian analysis, we wish to construct the {\it a posteriori} probability distribution of 
parameters ${\bf q}=(q_1, ..., q_N)$, $P({\bf q} \vert \data ,\th )$,  an update from the {\it a priori} probability $P( {\bf q}  \vert \th )$ on the  theory space $\th$ of the parameters that is driven by the likelihood ${\cal L}( {\bf q} \vert \data , \th )\equiv P( \data \vert {\bf q} ,\th )$ of the data $\data$ given ${\bf q}$:
%----------------------
\begin{equation*}
P({\bf q} \vert \data ,\th )=  P( \data \vert {\bf q} ,\th )  P( {\bf q}  \vert \th ) /P(\data  \vert  \th )
\end{equation*}
%--------------------
The prior may include theoretical prejudice, information derived from other data, and, at the very least, the specific measure adopted for the parameters. 
 The {\it evidence}, $P(\data \vert \th )$, a single normalization, is also needed  to ensure the posterior integrates to unity. Its determination is generally computationally-intense if one integrates over all parameter space, but it may only be needed  at late stages of reduction, e.g. over 2D and 1D reduced parameter spaces.

We can insert various further conditional probabilities on the path to
the confidence limits on $r$ from the fully reduced data. Examples in the transition from multichannel timestreams are: to multifrequency maps; to component-separated maps;  to bandpowers of cosmic spectra; to cosmic and nuisance parameters; to $r$. 
It is feasible to skip over the reduction-to-bandpowers step for Spider-like experiments because the number of pixels required will allow us to do a direct leap from the maps. 

We express the Bayesian chain for the posterior in terms of an
information action ${\cal S}_{\rm I}({\bf q})$,  an energy-like (in
temperature units) Euclidean action function that includes the
likelihood and the prior:
\begin{eqnarray}
&& P({\bf q} \vert \data ,\th ) \equiv e^{- \ln P(\data  \vert  \th ) } e^{-{\cal S}_{\rm I}({\bf q}) }\, ,  \cr 
&& {\cal S}_{\rm I} = - \ln P( \data \vert {\bf q} ,\th ) - \ln P( {\bf q}  \vert \th )  \, , \cr
&& P(\data  \vert  \th )  = \int d^N{\bf q} e^{-{\cal S}_{\rm I}({\bf q}) } \, .
\end{eqnarray}
The more elements there are in the chain, the more additive contributions there are to this energy. 
The evidence enters like the partition function in statistical mechanics, and its log is the  (negative of) a free energy  (in dimensionless units).   
%-------------------------------------------

\subsubsection{Reduction to Maps and Other Matched Filterings}\label{sec:mapmatched}

Initially $\data$ is in the form of time-ordered information, ToI's, containing the time-ordered data, and, typically, many flags about the data quality. The first step in the chain is to create maps from these, with  ${\bf q} $ being the map data vector ${\bf \Delta}$, with components $\Delta_{cxp}$  labelled by frequency channel $c$, Stokes polarization index $x={T,Q,U,V}$ and  spatial pixel number $p=1, ..., N_{\rm pix}$. The Stokes parameters ${Q,U,V}$ are referred to a fixed polarization sky reference frame in real space. (Most experiments do not have simultaneous ${T,Q,U}$ and $V$ detectors.)

The solution of the parameter estimation problem in this case is a set
of (generalized) pixel means ${\bf \bar{\Delta}}$, and a noise
covariance matrix  $C_{{\rm N}} = \avrg{\delta {\bf \bar{\Delta}}
  \delta {\bf \bar{\Delta}}^\dagger}$, in terms of the noise vector
$\delta {\bf \bar{\Delta}} ={\bf \bar{\Delta}} -  {\bf
  {\Delta}}$. Henceforth, we do not use bold letters for the matrices, which are the most
often used entities. 
The
way one does this is to solve $d=d_{\rm op}(q)+n$, with the operator $d_{\rm op}(q) =
\varphi q$, a linear data model with amplitudes $q$ and templates
$\varphi$. Here $d$ represents the time ordered data. The templates
form an $N_{\rm t}\times N_{\rm pix}$ 
matrix,  where $N_{\rm t}$ and $N_{\rm pix}$ are  the total number of  digitized time
observations and pixels (from all maps) respectively. This is a large compression of the data, by of order
$N_{\rm t}/N_{\rm pix}$, done by projecting out elements of the time-streams that are incompatible with the templates $\varphi $. (That projected-out information is a fertile residual space for searching for the  signals of relevance for experimental systematics studies.)

Making maps in this way is just one example of matched-filter
processing of linear data models. The main ingredient is an optimal
filter $\psi$ constructed from the linear templates $\varphi$ with
weight $w_{n{\rm f}}$: 
\begin{eqnarray}
 && d=\varphi q+ n  ,  \quad n=d-\varphi \avrg{q \vert d}_{\rm f} \nonumber \\
 && \avrg{q \vert d}_{\rm f} = \psi  (d-\bar{n}) +w_{q{\rm f}}^{-1}w_{q{\rm i}}\bar{q}_{\rm i}  , \quad   \psi \equiv  w_{q{\rm f}}^{-1}  \varphi^\dagger w_{n{\rm f}}\,  , \label{eq:matchedfil} \\
&&w_{q {\rm f}} = \varphi^\dagger w_{n{\rm f}}\varphi +w_{q {\rm i}} , \, \nonumber  \\
 && \delta q  \equiv q-\avrg{q \vert d}_{\rm f}  ,\,  \nonumber \\
 &&  \ \delta n \equiv n-\bar{n}  , \ \bar{n} = \avrg{n}_{\rm f} ,
  \ \bar{q}_{\rm i} =\avrg{q}_{\rm i} , \, \nonumber \\
 && w_{n {\rm f}}^{-1} = \avrg{\delta n\delta n^\dagger}_{\rm f},  \ w_{q{\rm f}}^{-1} = \avrg{\delta q\delta q^\dagger}_{\rm f} , \ w_{q{\rm i}}^{-1} = \avrg{\delta q\delta q^\dagger}_{\rm i}\, . \label{eq:wnoise}
 \end{eqnarray}

The residual $n$ is a "generalized noise" that is unaccounted for in the $\varphi q$ template representation of the data vector. We have allowed for a non-zero mean $\avrg{n}_{\rm i}= \bar{n} $ of the residual (e.g., it does not vanish in  the  cosmic parameter estimation of \S~\ref{sec:quadfil}).  The weight $w_{n{\rm f}}$ is optimally related to the correlation in the noise fluctuations, eq.~\ref{eq:wnoise}, in the sense that it minimizes the  final correlation matrix of parameter errors $w_{q{\rm f}}^{-1}$. Other weight choices than this optimized $w_{n{\rm f}}$ can work well, at the expense of  enhanced errors on the $q$ estimators. We have  added an initial  signal weight $w_{q{\rm i}}$, which is updated by the data to $w_{q{\rm f}}$. A $w_{q{\rm i}}$ is necessary if the dimension of the signal space exceeds that of the data space. If this is not the case,  we often  operate in the $w_{q{\rm i}}\rightarrow 0$ limit.  

If the residual variance $\delta n\delta n^\dagger$ is determined from the data $d$ itself, it would not be invertible. The estimation of $w_{n{\rm f}}$ requires an assumption to regularize the inversion, e.g., the raw variance is smoothed. The prior on  the form of $ w_{n{\rm f}}$ may turn it into an extra parameter estimation problem.  The compression of the $N_{\rm t} \times N_{\rm t}$ information in the matrix $\delta n\delta n^\dagger$ onto the parametric form can regularize $ w_{n{\rm f}}$. 
 
The derivation of  eq.~\ref{eq:matchedfil}  is most easily seen if
both the data noise $n$ and the signal prior for $q$ are Gaussian: 
\begin{eqnarray}
{\cal S}_{\rm I} &=& \half (n-\bar{n})^\dagger w_{n{\rm f}}^{-1}  (n-\bar{n}) 
+  \half N_d \ln (2\pi) +\half \Trace \ln w_{n{\rm f}}^{-1} \cr
&+& \half (q-\bar{q}_{\rm i})^\dagger w_{q{\rm i}}^{-1}  (q-\bar{q}_{\rm i})  
 +  \half N_q \ln (2\pi) +\half \Trace \ln w_{q{\rm i}}^{-1}  \, \nonumber , 
\end{eqnarray}
with $d=\varphi q+ n$. 
Manipulating this gives the usual:
\begin{eqnarray}
&& {\cal S}_{\rm I} ={\cal S}_{{\rm I}q} + {\cal S}_{{\rm I}d} \, ,  \cr
&&{\cal S}_{{\rm I}q} = \half  \delta q^\dagger w_{q{\rm f}} \delta q +    \half N_q \ln (2\pi) +\half \Trace \ln w_{q{\rm f}}^{-1} \, , \label{eq:matchedfil2}\cr
&&  {\cal S}_{{\rm I}d} = \half d^\dagger C_{\rm t}^{-1} d + \half N_d \ln (2\pi) +\half \Trace \ln C_{\rm t} \, , \label{eq:likefil}\cr
 && C_{\rm t} \equiv w_n^{-1}+\varphi w_{q{\rm i}}^{-1}\varphi^\dagger \nonumber  \,  . 
\end{eqnarray}

From ${\cal S}_{{\rm I}q}$, Wiener-filtered linear signals $\avrg{q \vert
  d,\bar{q}_{\rm i}}_{\rm f}$ (eq.\ref{eq:matchedfil}) and the fluctuations about
them, $w_{q{\rm f}}^{-1/2} \eta$, with $\eta$ an $N_q$ vector of Gaussian
random deviates, are obtained. Either this method, or approximations
to it,  is the preferred one for $E$ and $B$ construction. The first
such separated polarization component maps derived from data were
presented in the CBI papers \citep{ree06}, of course with the $BB$ a
non-detection consistent with the noise. There has been much
discussion about using variants of this approach for $E$-$B$
separation (e.g. \cite{lew02,bun02,bun03,bun11}).

From ${\cal S}_{{\rm I}d}$, the statistics of the cosmic (and other) parameters the $w_{q{\rm i}}$ in the prior depends upon are derived. This is our main focus here. 
%-------------------------------------------

\subsubsection{From Pixel Maps to $E$ and $B$ Maps}\label{sec:EBmaps} 

The map data vector ${\bf \Delta}$ is composed of a number of signals
${\bf s}$ as well as the map noise ${\bf n}$. The noise encompasses
true instrumental noise, experimental systematic effects, and
possibly, may draw terms from the signal side that are unwanted
residuals on the sky, e.g., from foreground subtraction
uncertainties. Each signal has a frequency dependence and polarization
components,  labelled by the Stokes parameter index $x$. The map
components are generally considered to be linear in the sky signals, 
\begin{eqnarray}
&& \Delta_{cxp}=\sum_J s_{Jc xp} +n_{cxp}, \quad \ x\in \{T,Q,U,V\},  \cr
&& s_{Jcx p} = \sum_{\ell m} \int_{\nu} {\cal F}_{cxp,J\nu x \ell m}   a_{J \nu x\ell m} , 
\end{eqnarray}
where the spherical harmonic signal amplitude for signal $J$ is  $a_{J \nu x\ell m}$. The transformation from this natural multipole space for the signals to the map space is encoded in the filters ${\cal F}_{cxp,J\nu x \ell m} $, which includes beam and pixelization information, the frequency response function for the channels, and a mask $\mu_{p \ell m}$. The mask $\mu$  could be a sharp cookie cutter or be more gently  tapered.

The $a_{J \nu x\ell m}$ are the coefficients in the standard expansion
of the CMB temperature and polarization fields in orthogonal mode
functions,  which are the spherical harmonics,  spin-$0$ for T and
spin-$2$  for polarization, with further linear combinations of the
spin-$2$ expansion coefficients defining the E and B modes:
 \begin{eqnarray}
&& T_{J \nu}(\theta,\phi)=\sum_{\ell=2}^{\infty} \sum_{m=-\ell }^{\ell }a_{J \nu T\ell m} Y_{\ell m}(\theta,\phi), \nonumber \cr
&& (Q\pm iU)_{J \nu}(\theta,\phi)=\sum_{\ell=2}^{\infty} \sum_{m=-\ell}^{\ell }  {}_{\pm2}a_{J \nu\ell m}~[_{\pm2}Y_{\ell m} (\theta,\phi)]\, , \label{eq:alm} \cr
&& a_{J \nu E\ell m}=-\frac{1}{2} ({}_2a_{J \nu \ell m}+{}_{-2}a_{J \nu \ell m})\ , \nonumber \cr 
&& a_{J\nu B\ell m}=-\frac{1}{2i}({}_2a_{J \nu\ell m}-{}_{-2}a_{J \nu\ell m})\, . \nonumber
\end{eqnarray}
The separation of the polarization into $E$ and $B$-modes is useful
because scalar perturbations only result in the $E$ mode whereas the
tensor  perturbations generate both (\cite{kam97},
\cite{zal97}). Nonlinear transport effects associated with the weak
lensing of the primary CMB fluctuations turn some scalar E mode into
scalar B mode, mostly at higher $\ell$s than the tensor component gives, so
separation for $r$ detection can be done. Note that this lensing
source has non-Gaussian features which means the power spectra are not
enough to characterize that signal. 

For Thompson scattering anisotropies, the $V$ Stokes parameter associated with circular polarization vanishes, as it also does for most Galactic foregrounds contaminating the primary CMB signal, so we now drop it from our consideration. It would of course be of interest to show experimentally that there is indeed no circular polarization in the CMB data. 

As we have noted above, eqs,~(\ref{eq:matchedfil}) and
(\ref{eq:matchedfil2}) can be applied to the case in which the $d$ are
the maps $\Delta_{cxp}$, the templates $\varphi$ are the $E$ and $B$
mode function rotators and the parameters $q$ are the $E$ and $B$
amplitudes in $\ell m$ space. Since these compressed maps $q$ and
their variances contain complete statistical information for a
Gaussian model, the $q$-power can be estimated from $\avrg{q \vert d}
\avrg{q \vert d}^\dagger + \avrg{\delta q\delta q^\dagger}$. This is
not the optimal determination of power. We adopt the CBIpol
approach \citep{sie04}, that while such optimal $E,B$ separation is good for checking robustness of results and for visualization of the polarization signals, the path to parameters (including bandpowers) is through the quadratic matrix methods, the mLikely approach of CBIpol, using eq.\ref{eq:likefil}.   
%-------------------------------------------

\subsubsection{Maps to Parameters with Matrix-based Likelihoods}\label{sec:mLike}
 
 For  statistically isotropic signals there are generally six
 cross-spectra among the coefficients, 
\begin{eqnarray}
&&\langle a_{x\ell m} a_{x^\prime \ell^\prime m^\prime }^{ *} \rangle =C_{X\ell} \delta_{\ell \ell^\prime}\delta_{mm^\prime }, \ 
X=xx^\prime, \  \, \cr 
&& {\rm for} \ x \in \{ T,E,B\}, \    X \in \{ TT,EE,BB,TE,TB,EB\} \, .  \nonumber
\end{eqnarray}
Typically the $EB$ and $TB$ power vanish (theoretically anyway) and only four are needed. However, $EB$ and $TB$ may be kept for systematics monitoring. 
For statistically homogeneous and isotropic 3D Gaussian initial
conditions, the primary CMB T,Q,U are isotropic 2D Gaussian fields
whose probability distribution depends only upon the   power spectra
$C_{X\ell}$, or, equivalently the X-power per $ \ln (\ell +1/2) $, 
\begin{eqnarray}
&&{\cal C}_{X\ell}\equiv \frac {\ell (\ell+1)}{2\pi} C_{X\ell} \, . \nonumber
\end{eqnarray}

If there is no correlation between signal and noise, the components of
the total covariance matrix $C_{t, {cxp}{c^\prime x^\prime p^\prime}}$
are given by the sum
\begin{eqnarray}
&&  C_{\rm t}= C_{\rm N}\ + \sum_{JJ^\prime}  C_{{\rm S}, JJ^\prime}  \, , \,  
C_{{\rm N},{cxp}{c^\prime x^\prime p^\prime}}=\langle n_{cxp}n_{c^\prime x^\prime p^\prime}\rangle \, , \cr \nonumber
&& C_{{\rm S},Jcxp,J^\prime c^\prime x^\prime p^\prime}=\langle s_{Jcxp}s_{J^\prime c^\prime x^\prime p^\prime}\rangle \, . \nonumber
\end{eqnarray}

The goal of bandpower estimation is to radically-compress the map information onto $\ell$-band power amplitudes the $q^{X\beta }$, with templates  $\varphi$ of form ${\cal C}_{X\beta ,X\ell}$. With sufficiently fine $\ell$-space banding, this stage of compression can be relatively lossless, allowing the cosmic parameters to be derived accurately. The inter-band shape of these templates may be crafted to look like theoretically expected shapes, or could just be flat, which imposes no prior prejudice. Both approaches have been effectively used. Usually the $\beta$-shapes have been chosen to be sharply truncated with no overlap in $\ell$-space, but this is not at all necessary. 

With cut-sky maps, bands are coupled even though they would not be for full sky observations with statistically homogeneous noise. The optimal method for estimating 
power spectra in the general case is the computationally expensive brute-force maximum
likelihood (MLE) analysis (e.g. \cite{bon98}), which iteratively corrects a quadratic expression for deviations $\delta q^\beta$ of the various bandpowers  from their initial values $q_{0}^\beta$ until the maximum likelihood  $q_{m}^\beta$ is reached. The weight matrix $C_{\rm t}^{-1}({\bf q})$ is adjusted at each step, until it settles into $C_{\rm t}^{-1}({\bf q}_m)$. The weight enters in two ways, one is quadratically in the likelihood-curvature matrix (approximately the Fisher matrix) and the other is in the force that drives the relaxation of the parameters to $q_{m}^\beta$. It turns out that one can think of the quadratic expression as describing the action of a matched filter on the pixel-pixel pair data product, similar to the way linear filters acting on the data vector may be matched, as we show in \S~\ref{sec:quadfil}.

Matrix methods for bandpower estimation were used by Boomerang
\citep{ber00,ruh03} and in all CBI papers. If the cosmic parameter of interest is linear, like $r$, then it can be viewed as a single template big-band bandpower. Even with the fully nonlinear ${\cal C}_{X\ell}(q)$, the amplitudes $\delta q$ can be iteratively solved for using linear derivative templates, and, with convergence, the result is the same as a full nonlinear treatment gives. 
%-------------------------------------------

\subsubsection{Pseudo-$C_{X\ell}$ cf. Map-Matrix Methods}\label{sec:pseudoCL}

Several fast  sub-optimal
approximate  methods have been developed to make the bandpower computations less computationally intense than in the map-matrix method: e.g.,  pseudo-$C_\ell$
estimators \citep{han03,cho04},  
SPICE \citep{sza01}, MASTER \citep{hiv02} and Xfaster \citep{contaldi,roc10}. Pseudo-$C_\ell$'s are constructed by direct spherical
harmonic transform of the cut-sky maps, or more generally, taper-weighted CMB maps. The all-sky bandpower centred on a specific $\ell_\beta$, $q^{X\beta}$, is then related by an appropriate filtering which draws the pseudo-$C_{X\ell}$'s from a wide swath of $\ell$'s determined by a mask-defined coupling matrix into the desired $\ell_\beta$-band. In spite of this $\ell$-space mixing, extensive testing has shown these methods to be accurate  for temperature anisotropies for large pixel numbers where the matrix inversions of the iterated quadratic approach are prohibitively expensive computationally. They have also been applied effectively to polarization datasets such as Boomerang \citep{mon06,pia06}.  

 The pseudo-$C_{X\ell}$ for $X=EE,BB$  suffer from $E$-$B$ mixing in addition to the $\ell$-space mixing: 
the estimated $C_{BB\ell}$ receives contributions from both $E$
and $B$-modes. The contamination coming from the $E$-mode can be
removed from $C_{BB\ell}$ in the mean by having the estimators undergo a
de-biasing step. However, there is still an extra contribution to
the variance of estimators which is due to the dominance of the
relatively large $E$ signal mixed into the $B$ measurement.
This can limit the primordial gravitational wave detection to $r\approx 0.05$ for deep small sky
surveys (covering about $1\%$ of the sky) as shown by (\cite{cha05}).
\cite{lew02} show how to construct window functions that cleanly
separate the $E$ and $B$ modes in harmonic space for azimuthally
symmetric sky observations at the cost of some information loss due to
the boundary of the patch.
In another treatment of the $E$-$B$ mixing problem, \cite{bun03}  show
that the polarization maps can be  optimally decomposed into
three orthogonal components: pure $E$, pure $B$, and ambiguous
modes. The ambiguous modes receive a non-restorable contribution from both $E$ and $B$ signals,
and are dominated by $E$
signal, thus should be removed in $B$-mode analysis. Based on this
decomposition, a near-optimal {\em pure} pseudo-$C_\ell$ estimator was
proposed (\cite{smi06}) and developed (\cite{smi07}, \cite{gra09}) which ensures no 
$E$-$B$ mixing. Recently \cite{bun11} has given a more efficient recipe for decomposing polarization data into $E, B$ and ambiguous maps, although still along the lines of \cite{bun03}. 

It is clear that if the full map-likelihood analysis  can be done,
then it should be done, since relevant information is not being thrown
away. There are two drawbacks to the map-based approach. The first is that $C_{\rm t}$
should saturate all contributions to signal and noise since we are in
quest of a small, essentially perturbative, component associated with
$r$ whose values can be biased by the missing components. This could be challenging in the presence of complex filtering resulting from time-ordered data processing.
Also the
computational cost of the required large matrix  manipulations is
high compared to the suboptimal methods. The
matrix size  depends upon the fraction of sky covered and the
resolution. For example, for an experiment covering $25\%$ of the sky
analyzed at a Healpix resolution of $N_{\rm side}=64$, the sizes are
$35K\times 35K$ and we find the likelihood calculation takes about $5$
minutes on a node with $16$ Dual-Core Power $6$ CPU's at $4.7$ GHz
(and theoretically capable of doing $600$ GFLOPS/node). In practice, our matrices are smaller than
this since the quest for $r$  requires a relatively low resolution analysis and only a  few other parameters that are correlated with it need to be carried along, as we show here.  To include many more parameters standard Bayesian sampling algorithms such as MCMC and adaptive importance sampling (\cite{wra09}) can be used. If we need to cover small angular scales as well as large, the matrices become prohibitively large, and hybrid methods, with a map-based likelihood for large scales joined to an $\ell$-space-based likelihood for small scales,  are needed.
%-------------------------------------------

 \subsection{The Downward Flow of Shannon Entropy  from  Data Compression onto  Theory Subspaces}\label{sec:shannonS}

The Shannon entropy $S_{\rm f}$ of the final (posterior) probability distribution is an average  of the log of the local phase space volume $\avrg{\ln p_{\rm f}^{-1} }_{\rm f}$ over the
posterior probability distribution $p_{\rm f}$, and is considered to
provide an estimate of the total information content in the final
ensemble (see, e.g., \cite{MacKay}):
\begin{eqnarray}
&&
S_{\rm f} (\th \vert \data ) = -\int d^N{\bf q} p_{\rm f} \ln p_{\rm f} = \avrg{\ln
  P({\bf q} \vert \data ,\th )^{-1}}_{\rm f}  \cr
&& =  \avrg{{\cal S}_{\rm I}}_{\rm f} + \ln P(\data  \vert  \th ) \nonumber \cr
&&\avrg{{\cal S}_{\rm I}}_{\rm f} \equiv  \int d^{N}{\bf q}  e^{-{\cal S}_{\rm I}} {\cal S}_{\rm I} /\int d^{N}{\bf q}  e^{-{\cal S}_{\rm I}} .  \nonumber
\end{eqnarray}
The initial entropy is averaged over the initial ensemble: $S_{\rm
  i}\equiv \avrg{\ln P({\bf q} \vert \th )^{-1}}_{\rm i}$. For a
uniform prior over a volume $V_{q,{\rm i}}$ in ${\bf q}$-space, it is
$S_{\rm i} = \ln V_{q,{\rm i}}$. The final entropy can be thought of as having a contribution from  (the log of) an {\it effective} phase space volume, reduced  relative to the initial one because of the measurement, plus a term related to the average $\chi^2$ associated with the mean-squared-deviations of $q$, usually just  the number of degrees of freedom unless the model is a very poor representation of the information content of the data. 

It should not seem curious to say that the information entropy decreases as a result of measurements, but it may seem curious to word it as: the average information content decreases. That is because the fully random initial state has more information, in that the variables can take on a wider range of values. We think  the reduced post-experiment information content is of higher quality. What constitutes Quality in information is subjective of course. 

Consider the initial space of the  $\data $, the space of full time-ordered-information, replete with bolometer readouts, flagged glitches, housekeeping information, etc. The amount of information we begin with is therefore enormous. From this data  optimal maps are constructed with map parameters ${\bf q}$ having channel and Stokes as well as pixel indices (the $ \Delta_{cxp}$ defined above) which defines the $\th$-space for this leg of compression.  
As the iterations progress towards the maximum likelihood map, there
is a mismatch between the noise power spectrum $w_{n{\rm f}}^{-1}$ on the
prior iteration, and the noise variance on the posterior iteration:
the latter will be less, hence so will its logarithm, hence so will
the Shannon information, until it settles into its final  converged
value, the information entropy in the maps, $S_f({\rm maps})$. Thus,
$S({\rm maps})$  decreases substantially from the large available
information in the uninformed prior, but also decreases as the
iterations converge, settling on $S_f({\rm maps})=N_{D}/2 +N_{D}\ln (2\pi)/2 + \Trace (\ln C_{\rm N})/2$. The new dimension for the reduced data  is the number $N_D$ of generalized pixels:  it is   the  total number of pixels from all channels times 3 for $T,Q,U$. The $N_D \times N_D$ noise matrix is $C_{\rm N}= (\varphi^\dagger w_{n{\rm f}} \varphi )^{-1}$. The information per generalized pixel is not so large but there are lots of such pixels. 

The standard noise assumption we make for our maps is  that it is
homogeneous and white, usually different for $T$ and $Q,U$. Given a
total integrated noise power, $N_{\rm pix}{\bar \sigma}_{\rm pix}^2$,
the map entropy is maximum if the noise is white with the same
${\bar{\sigma}}_{\rm pix}^2$ for each pixel. We have included modest
(yet realistic) inhomogeneity in the noise as well to test sensitivity
to this assumption, but find that makes little difference to our
results. The noise in a pixel of area $A_{\rm pix} $ from observations
covering an area $4\pi f_{\rm{sky}}$   over an  observing time
$T_{\rm obs}$ is $ \sigma_{\rm pix}^2 \propto 4\pi f_{\rm{sky}}/(A_{\rm
  pix} T_{\rm obs})$.  In that case, the entropy is $S({\rm maps})/pixel =
[1+ \ln (2\pi \sigma_{\rm pix}^2)]/2$. With a fixed $T_{\rm obs}$ and pixel size, the total entropy difference is $\propto f_{\rm{sky}} \ln f_{\rm{sky}}$  times a large number, thus quite a bit higher for large regions, and not just because there are more pixels: it is higher per pixel. The entropy in the map is more constrained if we focus our available resources on smaller regions, but of course only if the regions are of a size and resolution to be of relevance for our target cosmological parameter, e.g., $r$. 

We can obviously use the maps rather than the ToI's as our starting
point since, by design, no information relevant to estimation of our
target $r$  is lost in the compression. The pixel sizes are chosen so
this is true. Most of the huge entropy store in the ToIs is
inaccessible to $r$. As we have discussed, the traditional approach is
to  further compress $\data $, but in $\data \otimes \data $ space
(actually in the symmetric $\data \vee \data $ space), by solving for
bandpowers in the manner described above. The translation of the
variables is: $d$ are  now the map products ${\bf \Delta \Delta}^\dagger$, $\varphi$ is $C_{{\rm S},pp^\prime}$ and $\bf{{\bar q}}$ is the vector of (normalized) bandpowers. Because the bandpower likelihood surface $p_{\rm f} (\bf{{\bar q}})$ is quite complex, non-Gaussian and with band-to-band correlations, determining the information in the bandpowers requires a direct integration. As well, the $N_{{\rm band}}$-bands are generalized ones, indexed by channel number, polarization component (a number for $TT,TE,EE, BB, TB, EB$), as well as by $\ell$-band number.   With maximum likelihood relaxations to the bandpowers and Fisher matrix determination of errors (such as is used in XFaster), we would get $S({\rm band}) = N_{{\rm band}}/2 +  N_{\rm band} \ln (2\pi)/2 +  \Trace (\ln F_{\rm band}^{-1})/2$. 

An oft-used approximation to likelihood surfaces fully determines
$P(q^\beta)$ for each band $\beta$ with amplitude $q^\beta$, but
treats band-band correlations in a weakly coupled Gaussian
approximation. For example,  Boomerang and CBI and other CMB
likelihood analyses used the offset log-normal approximation of
\cite{bon98}:  each $P(q^\beta)$ was fit by a  Gaussian in the variable
$z^\beta=\ln (q^\beta + q_{\rm N}^\beta )$ which required an estimate of
the noise in the band $q_{\rm N}^\beta$ as well as the observational mean
$\bar{q}^\beta$, with a posterior of form 
\begin{equation}
-\ln P({\bar q} \vert \data , \th ) = \half {\bf \delta z}^\dagger{\cal F}_z{\bf \delta z} + \half N \ln (2\pi )  + \half \Trace \ln {\cal F}_z^{-1}, \nonumber
\end{equation}

in terms of the fluctuation  $\delta z = {\bf z-\bar{z}}$ about the
observational $z$-average $\bar{z}^\beta=\ln (\bar{q}^\beta +
q_{N}^\beta )$. The transformed correlation matrix is ${\cal F}_z^{-1} = \avrg{ {\bf \delta z} {\bf \delta z}^\dagger}$. For WMAP, a correction to this treatment was used, and for Planck a much more accurate characterization of the likelihood surface is needed, and continues to be under active development (e.g., \cite{roc10}).  For both, the  likelihood is a hybrid,  map-based  for the low $\ell$'s, and bandpower-based (with $\Delta \ell_\beta =1$) for  high $\ell$'s. 

A fully-characterized bandpower likelihood surface can of course be
used for $r$ estimation provided it is lossless. If only a few bands
$\beta$ are used, we can use intra-band template shapes with
amplitudes $r_{X\beta}$, which are approximately lossless for $r$; a
2-band calculation is shown in \S~\ref{sec:rXbeta}. Mostly we quote
single-band results, the one-step leap to $r$ from the maps, using
full-matrix posteriors, as we explore how different expenditures of
observational time for various experimental sensitivities lead to
changes in  the error. We primarily quote $2\sigma_r$ as our error
figure of merit, determined as explained in \S~\ref{sec:2sig}.

A better figure of merit than $2\sigma_r$ is the change in 1D Shannon entropy which tells us the average amount by which the log of the allowed volume in the $r$ parameter space shrinks in response to varying the experimental setups. 
It is 1D because we marginalize over all other $N-1$ parameters, the cosmic ones of interest and any nuisance parameters deemed necessary for the analysis, such as those characterizing uncertainties in calibration, beams,  bolometer $T$-$Q$-$U$ leakage, and foreground uncertainties. The final 1D {\it a posteriori} probability 
$p_{\rm f}(r\vert \data, \th_r )dr =\avrg{ \delta(r_{\rm op}-r) }_{\rm f} {\rm d} r = \exp[-{\cal S}_{1{\rm I}}(r)-\ln P(\data, \th )]{\rm d}r$ involves a 1D information action ${\cal S}_{1{\rm I}}(r)$, the integration over all parameters except the operator $r_{\rm op}$ whose value is constrained to be fixed at $r$. Here $\data$ refers to data, e.g., the maps, $\th$ refers to the overall theoretical framework, e.g., inflation-inspired tilted $\Lambda$CDM adiabatic  with the usual basic six cosmic parameters plus $r$, and 
$\th_r$ refers to $\th$ with  the $r_{\rm op}=r$ constraint. 

The 1D Shannon information entropy, $ S_{1{\rm f}}(r) = \avrg{{\cal
    S}_{1{\rm I}}(r)}_{\rm f} + \ln P(\data, \th ) $ ,  is best done by numerical
integration over the $r$-grid. The result is very simple if we
truncate the ensemble-averaged expansion of ${\cal S}_{1{\rm I}}(r)$ at
quadratic order:  
\begin{equation}
S_{1{\rm f}}(r) \approx \half +\half \ln (2\pi) +\ln (\sigma_r) =\half + \ln V_r , \nonumber
\end{equation}
 where $V_r$ (defined by the equation) is the compressed phase space volume for $r$ after the measurements. 

Although we have used the natural log to make the entropy expressions familiar for physicists, in information theory one often uses the binary logarithm, $lb\equiv log_2$. With natural logs the information is in {\it nats}, but with $lb$ it is in {\it bits}. When expressing information differences in \S~\ref{sec:concl} we translate to bits. Since a full bit represents a factor of 2 improvement in the error bar, $\Delta S_{1{\rm f}}(r)$ may only be a fraction of a bit, trivial perhaps, but subtle too,  given the mammoth information compression from raw data to this one targeted parameter degree of freedom.

%------------------------------------------------
\subsection{$2\sigma$ Calculation}\label{sec:2sig}
 
We define $\sigma_{95}$ through :
\begin{equation}
\int_{\max(0,r_{\rm b}-\sigma_{95})}^{r_{\rm b}+\sigma_{95}} {\cal
  L}(r) {\rm d}r =0.954 \int_0^\infty  {\cal L}(r) {\rm d}r
\label{eq:2sig}
\end{equation}
where $r_{\rm b}$ is the best-fit value of $r$. The $\sigma_{95}$-limit is determined by numerically integrating the Gaussian-fitted $1$D
likelihood curve. 

In most cases considered in this paper the likelihood curves turn out to be well
approximated by Gaussians.  Therefore, when there is a few $\sigma$ detection (e.g. for
$r=0.12$) or  when $r\sim 0$, to a very good approximation
$\sigma_{95}=2\sigma$ where $\sigma$ is the width of the Gaussian
fit. Thus, throughout this paper we will use the common notation of
$2\sigma$ which represents $\sigma_{95}$ and has been calculated
through eq.~\ref{eq:2sig}. The only exception to this way of determining $2\sigma$
is when it is being directly given by the inverse of the Fisher
matrix, where $\sigma$ represents the width of the likelihood function, under
the assumption of its Gaussianity.    

%=========================================================
\section{Constrained Correlations and Linear Response In Pixel-Pair and Parameter Space}\label{sec:LRT}
 %-------------------------------------------

  \subsection{Matched Filters in Quadratic Pixel-Pair Space and Maximum Likelihood Estimation}\label{sec:quadfil}

When $C_{\rm t}({\bf q})= C_{\rm N} + C_{\rm S}({\bf q}) $ depends in a nonlinear way on ${\bf q}$, we can still explore the posterior space by a sequence of linearized steps  $\delta q^\alpha$ which converge to zero 
in the approach to the maximum likelihood;  $C_{{\rm t}*}=C_{\rm t}({\bf q}_*)$
evaluated at the prior step $q_*$ can be thought of as the new general
noise matrix  and $\partial C_{\rm S}({\bf q})/\partial q^\alpha \delta
q^\alpha $ the new signal matrix in the  linear model. The quadratic
expression determining the step is the action of a matched filter on
the pixel-pixel pair data $d \equiv {\bf \Delta\Delta}^\dagger$, and has a form that can be unravelled from the general expression 
 eq.~\ref{eq:matchedfil}, with a non-zero residual mean $\avrg{n}=C_{{\rm t}*}$. If instead of the value at the last iteration we take ${\bf q}_* =0$, we get the usual map noise $C_{\rm N}$, and a generalized noise with some signal contribution to it if  only some of the ${\bf q}_* $ are non-zero (e.g., foreground residual parameters).  
 
The signal coefficients $q_{X\beta}$ would be the isotropic power spectra bandpowers for the sets $TT$, $EE$, $BB$, $TE$, $TB$, $EB$. If the bands are of width $\Delta \ell =1$ consisting of a single multipole, but all $m$, the $\varphi$ are the filters for defining the pixel-pixel correlation matrices in terms of an  $\ell, m$ expansion of the total and polarization fluctuations, expressed in terms of the filters of \S~\ref{sec:EBmaps}, $\sim \sum_m {\cal F}_{cxp,J\nu x\ell m}{\cal F}^*_{cx^\prime p,J\nu x^\prime\ell m}$. Or we can choose just one $\ell$ band with a  template shape for each of the 6 $X$ cases with 6 amplitudes $r_X$ multiplying these. An example of this approach  is shown in \S~\ref{sec:rXbeta}. Or we could choose just one set of shapes for all 6, with only one amplitude multiplier, $q$ which we can normalize to be $r$. Template consistency is therefore assumed, and this gives the maximum leverage for teasing out the best  determination for $r$ from the data, although it is of course heavily conditioned by the assumptions that go into the template construction (namely the values assumed for the other cosmological parameters which fix the structure of the templates). 

The pixel-pair  residual fluctuation weight, ${\cal W}\equiv w_{n{\rm f}}=
\avrg{\delta n\delta n^\dagger}^{-1} $ is, for Gaussian models of
${\bf \Delta}$, expressible as quadratic combinations of
$w=\avrg{n}^{-1}=C^{-1}_{t*}$: 
\begin{equation*}
{\cal W}_{(ij)(kl)} =[w_{ik}w_{jl}+ w_{jk}w_{il} +w_{ij}w_{kl}]/4 . 
\end{equation*}
The inverse is  
\begin{equation*}{\cal W}^{-1}_{(ij)(kl)} =C_{t*,ik}C_{t*,jl}+ C_{t*,jk}C_{t*,il} +C_{t*,ij}C_{t*,kl}, 
\end{equation*}
 related so that  
 \begin{equation*}
 {\cal W}_{(ij)(kl)} {\cal W}^{-1}_{(kl)(mn)}=\delta_{(ij)(mn)} \equiv (\delta_{im} \delta_{jn} + \delta_{jm} \delta_{in})/2.
 \end{equation*}
(We use the Einstein summation convention, that like indices are to be summed.) 
 When we reorganize the $\varphi^\dagger {\cal W}$ projector on the right hand side and the $\varphi^\dagger  {\cal W}\varphi$ inverse residual matrix on the left hand side, we obtaine the familiar Fisher expression for the parameter response $\delta q^\alpha$  driven by the  
 pixel-pair deviation $\delta C_{\rm tO} \equiv {\bf \Delta \Delta}^\dagger-C_{{\rm t}*}$  
of the raw observational correlation function $C_{\rm tO}$  from its
current estimate $C_{{\rm t}*}$:
\begin{eqnarray}
F_{\alpha \beta}\avrg{\delta q^\beta \vert \delta C_{\rm tO} }&& =\half
\Trace [ C_{{\rm t}*}^{-1} \partial C_{\rm t} /\partial q^{\alpha}
  C_{{\rm t}*}^{-1}({\bf \Delta \Delta}^\dagger-C_{{\rm t}*}) ]  \cr
&& = [\varphi^\dagger {\cal W}({\bf \Delta \Delta}^\dagger-C_{{\rm t}*})]_\alpha \, , \cr 
 F_{\alpha \beta} && =\half \Trace [ C_{{\rm t}*}^{-1} \partial C_{\rm t} /\partial q^{\alpha} C_{{\rm t}*}^{-1}\partial C_{\rm t} /\partial q^{\beta}]  \cr
 && = [\varphi^\dagger  {\cal W}\varphi]_{\alpha \beta} \, . \label{eq:Fab}
\end{eqnarray}
  This expression shows that the $\delta q^\alpha$ -adjustment is through a matched filter based on the templates $\varphi^{X\alpha}_{X^\prime{(ij)}}$ of form $[\partial C_{\rm t} /\partial q^{X\alpha}]_{X^\prime (ij)}$. The weighting in pixel-pair space shown is essential for it to be optimal.  

In \S~\ref{sec:cXellLRT} and \ref{sec:rLRT}, we replace  $\delta C_{\rm tO} $ by other pixel-pair deviations to show how the single tensor template-based bandpower, namely $r$, responds to individual $E$ and $B$ multipoles - i.e., a window function showing where the $\ell$ power that $r$ is sensitive to lies. With the noise and sky fraction embedded in the weights and in $\varphi$, these window functions vary from experimental setup to experimental setup. 

%-------------------------------------------
\begin{figure*}
\begin{center}
\includegraphics[scale=0.8]{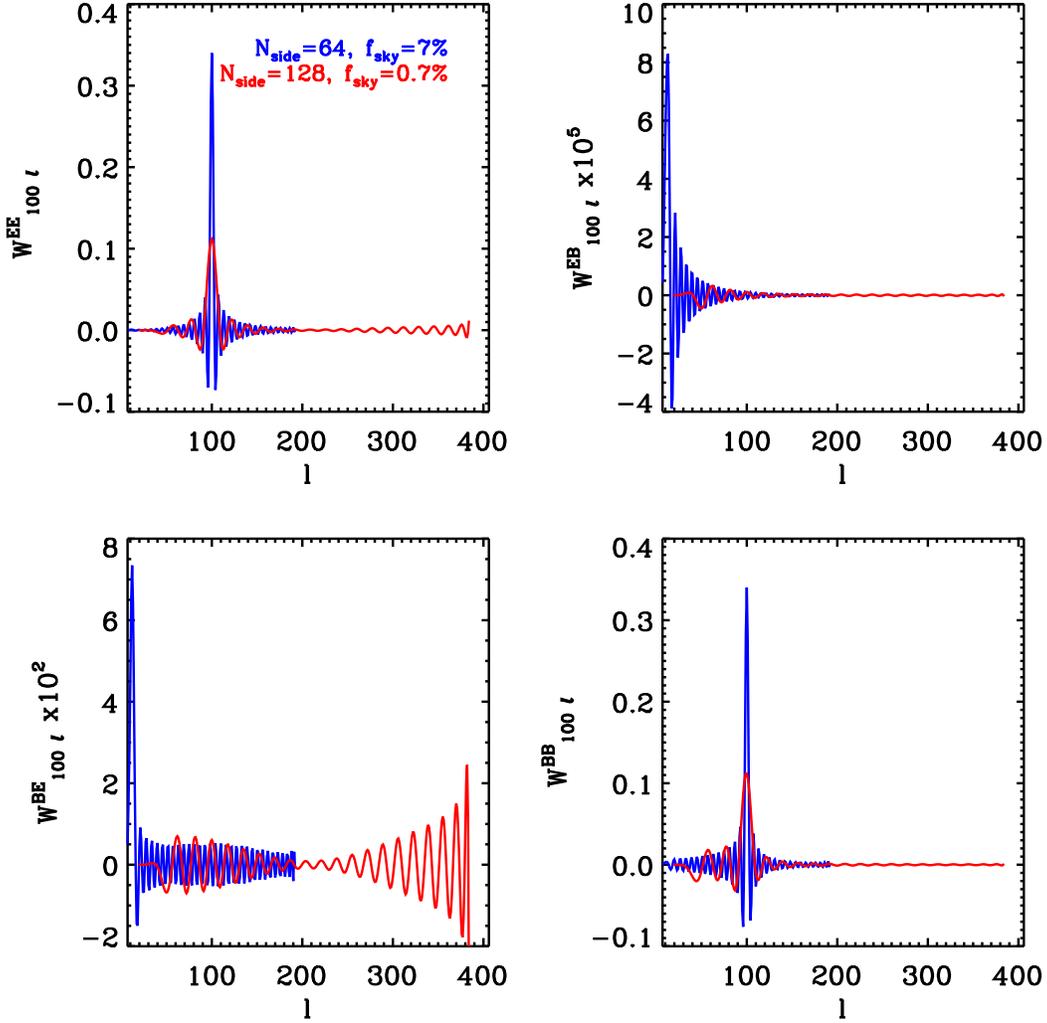}
\end{center}
\caption{The filter $W_{X\ell X^\prime \ell^\prime}$, $X,X^\prime \subset \{ EE,BB \}$, shows how  the mode 
   ${\cal C}_{X\ell}$ linearly responds to a small change in the mode ${\cal C}_{X^\prime \ell^\prime}$. The leakage response shown here is for an $\ell^\prime=100$ stimulus,  for a Spider-like experiment with 
    $f_{\rm{sky}}=0.07$ (at $N_{\rm side}=64$) and $f_{\rm{sky}}=0.007$ (at $N_{\rm side}=128$).  Note the different $y$-axis
  scales. }
\label{filter_res6_7}
\end{figure*}
%-------------------------------------------
%------------------------------------------
\begin{figure}
\begin{center}
\includegraphics[scale=0.45]{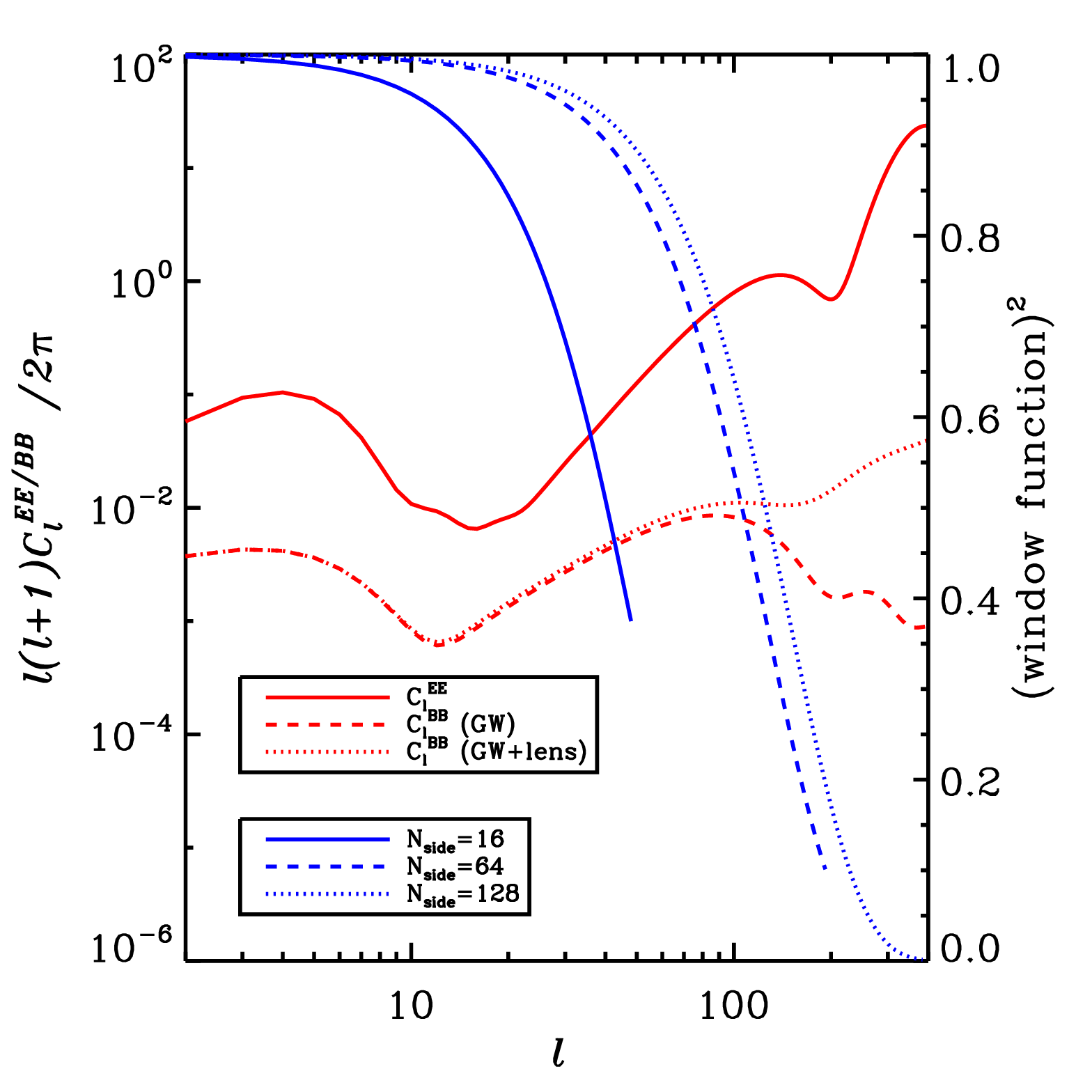}
\end{center}
\caption{Beam and pixel window functions for
  different resolutions are compared to the polarization power
  spectra for the best fit WMAP7-only parameters for the $\Lambda$CDM
  + lensing + SZ + tensor model, with the addition of a tensor
  component of strength $r_{\rm fid}=0.12$. 
   $B$-mode (GW) shows just the 
  gravitational wave induced contribution and  $B$-mode (GW+lens) includes the lensing contribution as well.}
\label{window}
\end{figure}
%-------------------------------------------

Equation~\ref{eq:Fab} is an exact one following from a $\chi^2$
minimization of the linear expansion of $C_{\rm t}$ in $\delta q$, albeit to
be iteratively corrected. The more data-related path is to expand the
information action associated with the posterior $p_{\rm f} ({\bf q}+\delta
{\bf q})$ about the  starting point  ${\bf q}_*$  to second order in
$\delta q^\alpha$ (e.g.,  \cite{bon96,bon98}),   
\begin{eqnarray}
&& {\cal S}_{\rm I} ({\bf q}_*+\delta {\bf q})={\cal S}_{\rm I} ({\bf q}_*) - p_\alpha \delta q^\alpha +\frac{1}{2} {\cal F}_{\alpha \beta}  \delta q^\alpha \delta q^\beta \, ,  \nonumber \cr
&& p_\alpha ({\bf q}_*)  \equiv - \frac{\partial {\cal S}_{\rm I} }{\partial q^\alpha} \, , \  {\cal F}_{\alpha \beta}({\bf q}_*) \equiv \frac{\partial^2 {\cal S}_{\rm I} }{\partial q^\alpha \partial q^\beta}, \nonumber  \cr
&& p_\alpha ({\bf q}_*) = \half \Trace [ C_{{\rm t}*}^{-1}\frac{\partial C_{\rm t}}{\partial q^{\alpha}} C_{{\rm t}*}^{-1}\delta C_{\rm tO}] \, , \nonumber \cr
&&  {\cal F}_{\alpha \beta}({\bf q}_*) -  F_{\alpha \beta}({\bf q}_*) =- \half \Trace C_{{\rm t}*}^{-1}\frac{\partial^2  C_{\rm t}}{\partial q^{\alpha}q^{\beta}}C_{{\rm t}*}^{-1}\delta C_{\rm tO}  \nonumber \cr 
&&+  \Trace C_{{\rm t}*}^{-1}\frac{\partial C_{\rm t}}{\partial q^{\alpha}} C_{{\rm t}*}^{-1}\frac{\partial C_{\rm t}}{\partial q^{\beta}} C_{{\rm t}*}^{-1}\delta C_{\rm tO} \, . \label{eq:calFab}
\end{eqnarray}
The fluctuation ${\cal F}-F$ of the curvature metric $ {\cal F}_{\alpha \beta}$ from the Fisher matrix $ F_{\alpha \beta}$ of eq.(\ref{eq:Fab}) has the two terms shown. Both are associated with the  residual $\delta C_{\rm tO} $ mismatch since the parameter space correlations may not be able to fully saturate the data correlations. If the theory (including noise) is a good approximation to those components of $\delta C_{\rm tO} $ which survive the heavy matched-filtering,  then these  terms disappear with ensemble-averaging over all realizations.  A caution is of course that we only inhabit a single realization. (The first subdominant second order term depends upon  $\partial  C_{\rm t} /\partial q^{\alpha} \partial q^{\beta}$, hence vanishes in a linear expansion model.) Each $\delta q^\alpha = [{\cal F}^{-1}]^{\alpha \beta} p_\beta$ drives the system towards the $p_\alpha ({\bf q}_*+\delta {\bf q}) =0$ "equilibrium", but corrective steps are needed to fully relax to $q_m^\alpha$.  In practice, using $F_{\alpha \beta}$ in place of ${\cal F}_{\alpha \beta}$ is usually adequate, and indeed often preferred.  In cases with structure-less likelihood functions, a few iterations usually suffice to take us as close to the peak as required.

Since the entire statistics, given the validity of the Gaussian approximation for both signal and noise, is fully specified by the likelihood expression together with the prior probability defining the measure on the parameter space, no issue explicitly arises about mixing the $EB$-modes.  The optimal quadratic filter to obtain the maximum likelihood for $r$ takes into account all aspects of the polarization. We can operate in the $QU$ polarization space, with specific spatial axes chosen for the polarization basis vectors, or we can do a transformation to spherical harmonic space and choose a polarization basis which is explicitly  $\ell m$ dependent, as in the $EB$ basis case. 
%-------------------------------------------

\subsection{ Linear Response of $C_{BB\ell}$ to $C_{EE\ell}$: Power Leakage}\label{sec:cXellLRT}

In this section, we use  quadratic matched-filters to quantify the leakage of CMB power
among the different ${\cal C}_{X\ell}$ spectra. These are
"susceptibilities", relating the linear response of a target variable
to the stimulus of a driver variable. We also refer to them as window
functions to be consistent with the language used for bandpowers, in
which the driver is the ${\cal C}_{S, X\ell}$, and the response is the
bandpower. The window function attached to each bandpower  "gathers in
$\ell$-space" from a given ${\cal C}_{S,X\ell}$ the bandpower. There
is a  long history of making such windows publicly available. They
were used in likelihood evaluations in the 2000 release of the Boomerang "B98" results \citep{lan01}. \cite{teg01} used similar window functions in the quest for a best  quadratic estimator.

If we treat ${\cal C}_{{\rm S},X\ell}$ as our variable and replace $C_{{\rm t}*}^{-1}\delta C_{\rm tO}$ 
by its ensemble average, we have
\begin{equation}\label{isotropy}
\avrg{\delta C_{\rm tO}} = \delta C_{\rm t} =\sum_{X,\ell} C_{{\rm S}pp^\prime ,{\cal C}_{X\ell}}\delta
{\cal C}_{X\ell} \, , 
\end{equation}
hence 
\begin{eqnarray}
&& \frac{\delta q^\alpha}{q^\alpha}=\sum_{X,\ell} W^\alpha_{X\ell} \frac{\delta {\cal C}_{X\ell}}{{\cal C}_{X\ell}}, \cr
&& W^\alpha_{X\ell}=\frac{{\cal C}_{X\ell}}{q^\alpha} \sum_{\beta} [F^{-1}]^{\alpha \beta} F_{\beta {X\ell}} ,  \cr
&&  F_{\beta {X\ell}} \equiv \avrg{\partial^2 {\cal S}_{\rm I} /\partial q^\beta \partial {\cal C}_{X\ell}}.\nonumber
\end{eqnarray}
It is isotropized over $m$. 
Another  variant is $W_\alpha^{X\ell}$ which can tell us how uncertainty in $q^\alpha$
 is distributed over $\ell$-space.  

With the ${\cal C}_{X\ell}$ as the response parameters
$q^\alpha$ as well as the stimulating drivers, we have 

\begin{eqnarray}
&& \frac{\avrg{\delta {\cal C}_{X\ell} \vert \delta {\cal C}_{X^\prime \ell^\prime}}}{{\cal C}_{X\ell}} =
\sum_{X^\prime ,\ell^\prime}W_{X\ell X^\prime \ell^\prime} \frac{\delta {\cal C}_{X^\prime \ell^\prime}}{{\cal C}_{X^\prime \ell^\prime}},  \nonumber \cr
&& W_{X\ell X^\prime \ell^\prime}= \frac{{\cal C}_{tX^\prime \ell^\prime}}{{\cal C}_{tX\ell}} \sum_{X^{\prime\prime},\ell^{\prime\prime}}[F^{-1}]^{X\ell,X^{\prime\prime} \ell^{\prime\prime}} F_{X^{\prime\prime}\ell^{\prime\prime},X^{\prime}\ell^{\prime}}\, . \nonumber 
\end{eqnarray}

We have verified numerically that a full sky observation using the matrix methods gives 
uncorrelated modes $W_{X\ell X^\prime \ell^\prime} =\delta_{\ell\ell^\prime}\delta_{XX^\prime}$. 
Figure \ref{filter_res6_7} shows the increase in mode correlation with decreasing $f_{\rm{sky}}$ for a fixed observation time. The observed
patches are in the form of spherical caps. We plot  an  $\ell=100$ stimulus for $f_{\rm{sky}}=0.07$
(at $N_{\rm side}=64$, pixel size $\approx56'$) and
$f_{\rm{sky}}=0.007$ (at $N_{\rm side}=128$, pixel size
$\approx 28^\prime$).  
(Figure~\ref{window} shows the associated beam and pixel window
functions along with the polarization power spectra.)  

Although the  $EE$ and $BB$ responses are localized around the input $\ell=100$,  they are spread over $\ell$ and leak into the other $X$-mode.  By contrast, the cross-filters ($W_{BE,100 \, \ell}$ and
$W_{EB,100\,\ell}$) are 
not  localized. Note that they are substantially smaller than $W_{BB,100\, \ell}$ and $W_{EE, 100\, \ell}$.
 We also see that the relative contribution of the $EE$ signal to the
contamination of $BB$ is about $3$ orders of magnitude larger than the
contamination in $EE$ due to $BB$.  We can conclude that $EE$ power uncertainties from a large range of scales will affect the  $BB$
measurement. 
The width of the oscillation $\Delta \ell \sim
\theta_{\rm patch}^{-1}$ is related to the cap size, 
narrowing as $f_{\rm{sky}}$ goes up.  
The leakage is  larger for smaller $r$, hence must be well characterized for  highly sensitive 
$B$-mode experiments to avoid a false detection.
%-------------------------------------------
\begin{figure}
\begin{center}
\includegraphics[scale=0.7]{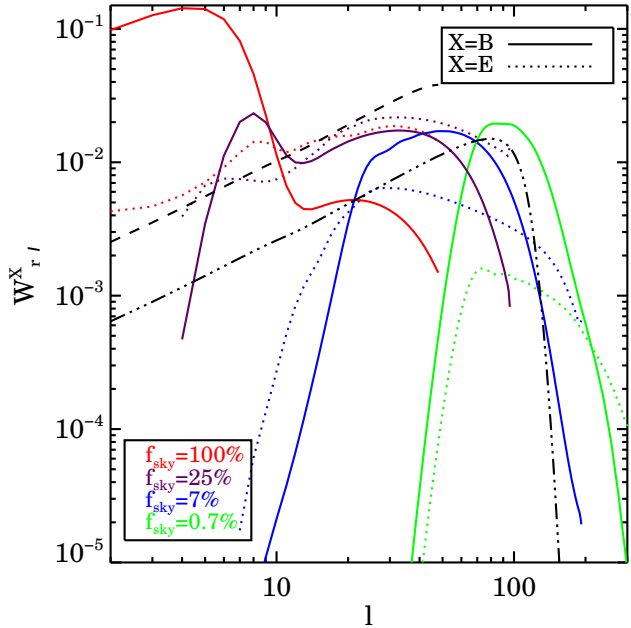}
%dlnr_dlncl.r.0.12.eps}
\end{center}
\caption{Window functions $W^{r}_{X\ell} $ for $X\subset \{ EE,BB \}$ for
  different sky cuts show that, as expected, all-sky experiments are nicely sensitive to the reionization $BB$ bump, but smaller sky experiments are not, although they pick up well the $\ell \sim 50-100$ region. We have used $r=0.12$ for the fiducial model. The rapid declines to high $\ell$ are more due to the onset of experimental noise than to the onset of the lensing-induced $B$ "noise". Residual foreground noise has not been included in these plots. Note that even a coverage with $f_{\rm{sky}}$ only 0.007 can punch out a robust detection from 50 to 150  in $\ell$, and though 0.07 loses out a bit (relatively) at 150, its detection would come from a wider stretch in $\ln \ell$, out to $\ell \sim 20$ before falling off. Only at $f_{\rm{sky}}> 0.25$ does one begin to pick up the reionization bump. The curious drop in  the all-sky $N_{\rm side}=16$ red line at the top is due to the Spider-like noise for higher $\ell$ being heavily enhanced because all of the sky is covered in the same amount of observing time. To illustrate the role of this, a CMBpol-like experiment with $C_{\rm N}$ decreased by $\sim 1000$  is plotted, with $N_{\rm side}=16$ (dashed straight line) and $N_{\rm side}=64$ (triple-dot-dashed line). The reason all three are offset from one another is because the normalizing  $\sigma_{r}^2$ depends upon the amount the filter captures of the total $r$ signal. }
\label{filter_r}
\end{figure}
%-----------------------------------------

%------------------------------------------

\subsection{Linear Response of $r$ to ${\cal C}_{BB\ell}$ and ${\cal C}_{EE\ell}$}\label{sec:rLRT}

We now use these quadratic matched-filters to quantify the linear
response of $r$ (and other cosmological parameters) to uncertainty in
the ${\cal C}_{X\ell}$, 
\begin{eqnarray*}
\frac{\avrg{\delta r \vert  {\cal C}_{X\ell}}}{r}  &=& \sum_{X,\ell}W_{r,X\ell}\frac{\delta {\cal
    C}_{X\ell}}{{\cal C}_{X\ell}}, \\
     W^{r}_{X\ell}&=&\frac{F_{rX\ell}}{F_{rr}} \frac{{\cal C}_{X\ell}}{r}.
\end{eqnarray*}
The filter for a Spider-like experiment with
a fiducial $r=0.12$ is shown in Figure \ref{filter_r},  as $f_{\rm{sky}}$ varies (as does the pixel size).  The red, purple, blue and green curves
correspond to $f_{\rm{sky}}=0.75, 0.25, 0.07$ and
$0.007$, calculated at $N_{\rm side}=16$, $N_{\rm side}=32$, $N_{\rm side}=64$ and
$N_{\rm side}=128$ respectively.
 As expected, the figures show that the measured $r$ is more sensitive to $BB$ than to $EE$
on most scales.

%=========================================================

\section{ Simulation Methods and Calculational Results}\label{sec:results}

In this section, we  use the map-based $TQU$ likelihood procedure of \S~\ref{sec:likelihood} to compute the posterior $P({\bf q}\vert f_{\rm{sky}}, \data , \th )$ in parameter subspaces and, by marginalization, the 1D posterior $P(r \vert f_{\rm{sky}},\data , \th )$ as a function of $f_{\rm{sky}}$.  Although we avoid  
 explicit $E$-$B$ decomposition with this method, we do make identical calculations to the $TQU$ matrix ones in $\ell$-space using $TT$, $TE$, $EE$ and $BB$, and assuming no mixing. We show that such a naive approach does quite well in predicting the errors: if properly handled, polarization-mode-mixing is not a significant error source in most cases. Of course for either method to be successful, all generalized noise sources need to be identified including instrumental leakage from $T$ to $Q,U$.  
%-------------------------------------------

\subsection{Calculation of Ensemble-Averaged Posteriors on Parameter Grids}\label{sec:ourMLE}

We calculate the posterior distribution on a gridded parameter space, a method mostly applicable to low
dimensional parameter spaces. At each point of the parameter grid the ${\cal C}_{X\ell}$'s are calculated
using the public code CAMB \footnote{http://camb.info/}. These are then multiplied by beam windows, ${\cal B}_\ell^2 =  e^{-\ell(\ell+1)\sigma_b^2}$, assuming a Gaussian beam of width $\sigma_b=0.425 \theta_{\rm FWHM}$,  and pixelization windows $W_{{\rm pix},\ell}^2$, an  isotropized approximation to finite pixel size effects. (Timestream digitization filters are also generally required, but are swamped by these two filters.) 
The product is used to construct the symmetric $3\times 3$ theoretical pixel-pixel signal covariance matrices, with 6 independent sub-matrices, $C_{{\rm S},X}$, $X\subset  \{ TT, TQ, TU,  QQ, QU, UU\}$.  We  assume experimental noise is Gaussian and usually take it to be white, so $C_{{\rm N},T}=\sigma_{n,T}^2 {\bf I}$  for the temperature block and $C_{{\rm N},Q,U}=\sigma_{n,{\rm pol}}^2 {\bf I}$ for the polarization block of the covariance
 matrix, where we usually have $\sigma_{n,{\rm pol}}\sim\sqrt{2}\sigma_{n,T}$. Here the $\sigma_n$ are effective noises per pixel, an amalgamation of the noises  coming from different frequency channels.  ${\bf I}$ is the identity matrix. We neglect leakage from $T$ to $Q,U$. 

Since we are forecasting the uncertainties in $r$ from different experimental setups, and not analyzing actual CMB maps, we can
bypass  creating a large ensemble of simulated CMB maps by replacing
the observed correlation matrix $C_{\rm tO}$  by its ensemble average:
\begin{eqnarray}
&&\Trace C_{\rm t}^{-1}{\bf \Delta\Delta}^\dagger  \rightarrow \Trace
  \avrg{C_{\rm t}^{-1}{\bf \Delta\Delta}^\dagger}=\Trace 
  C_{\rm t}^{-1}\bar{C}_{\rm tO}. \nonumber
\end{eqnarray}
 Here ${ \bar{C}}_{\rm tO}$ is the ensemble-averaged "pixel-pair data", namely the covariance matrix of the input
 fiducial signal model together with the instrument noise, and $C_{\rm t} ({\bf q})$ is 
 the signal pixel-pixel covariance matrix for the  parameters ${\bf q}$ plus the various noise contributions, instrumental and otherwise. 
An advantage of this approach is that the 
 recovered values of the parameters are what the ensemble average of
 sky realizations would yield, and will not move hugely due to the chance strangeness of any one realization (as the real sky may provide for us).
 Note that while sample variance does not impact the location of the maximum
 likelihood in this ensemble-averaged approach, it is fully reflected in the width of the posterior distribution from
 which our uncertainties are derived.

%---------------------------------------------
\begin{deluxetable*}{ccccccc}
\tablecaption{Specifications of Spider-like, Planck-like and CMBPol
  (mid-cost) experiments for simulations. }
\tablehead{\colhead{Experiment} & \colhead{Freq} & \colhead{FWHM}
  & \colhead{num. of det.} & \colhead{$\Delta T$\tablenotemark{a}} &
  \colhead{$\Delta T$} & \colhead{obs. time}\\
&(GHz)&&&$I$&\colhead{$Q~\&~ U$}&}
\startdata
Spider-like\tablenotemark{b}        & $96$   & $50'$ & $768$  & $3.2$ &$4.5$ & $580$ hr \\
Spider-like                         & $150$  & $32'$ & $960$  & $2.7$ &$3.8$ & $580$ hr \\
Planck-like\tablenotemark{c}        & $100$  & $10'$ & $8$    & $3.8$ &$6.1$ & $2.5$ yr \\
Planck-like                         & $143$  & $7'$  & $8$    & $2.4$ &$4.6$ & $2.5$ yr \\
CMBPol (mid-cost)\tablenotemark{d}  & $100$  & $8'$  & --     & $0.18$ &$0.26$ & -- \\
CMBPol (mid-cost)  & $150$ & $5'$   & --  &$0.19$ &$0.27$ & -- 
\enddata
\tablenotetext{a}{${\rm nK}$, the instrument sensitivity divided by
 $\sqrt{{\rm total ~ observation ~time}}$.}
\tablenotetext{b}{These Spider-like specifications which are used as the default in this paper are different from
  the more recently proposed ones in \cite{fra11} with 
  two $20$ day flights. The first flight  uses three
  $90$ and three $150$ GHz receivers each with $288$ and $512$
  detectors respectively. In the second flight, two $280$ GHz
  receivers replace one $90$ and one $150$ GHz telescope, leaving the
  configuration of the flight identical to the first one. 
  The detector sensitivity as proposed in   \cite{fra11} is $150$, $150$
  and $380 ~\mu {\rm K_{ CMB}}\sqrt{\rm s}$ at $90$, $150$ and $280$ GHz, respectively.
 The performance of the default Spider-like
  experiment in this paper and the more recent proposal as in
  \cite{fra11} are very close (see Figure ~\ref{fig:spider_apra_scip}).}
\tablenotetext{c}{http://www.rssd.esa.int/index.php?project=planck}
\tablenotetext{d}{For a mid-cost full-sky CMBPol experiment based on table 13 of
  \cite{bau09}. We are using $100$ and $150$ GHz channels in our
  simulations. Adding more channels, in the unrealistic case of no
  foreground contamination we simulate, would not affect the limits on
$r$, since with these low instrument noise levels, either lensing or cosmic
  variance, depending on how small $r$ is, would be the dominant
  source of uncertainty. }
\label{experiments}
\end{deluxetable*}
%----------------------------------------------
 
We mask out the part of Galaxy falling in the observed patch (the P06
WMAP-mask \cite{pag07}), assuming it to be too foreground-dominated
for useful parameter extraction. We also project out modes larger than
the fundamental mode of the observed patch since, due to time-domain filtering, information is not 
usually recoverable on such large scales.
 For instance, if the mask has the shape of a spherical cap
 extending from the north pole to $\theta=\theta_{\rm patch}$, we add a very
 large noise to the modes with $2\ell+1 < [2\pi/\pomega]$ where $[..]$ takes
   the integer part and 
 $\pomega=2\sin(\theta_{\rm patch}/2)$ is the flat 2D radius of the disk with an
 area equal to the solid angle of the cap. This makes the likelihood insensitive to
 any information at and beyond the patch scale. This large scale mode cut is especially
 important to include for larger values of $f_{\rm{sky}}$, where  the low
 $\ell$ modes contribute significantly to $r$ measurement through the reionization bump. In
 real large sky experiments it will not be easy to draw such modes
 from the maps. 

 Our simulations cover two observational cases: an all-sky experiment with 
 Planck-like white noise levels, and a partial sky experiment with
 Spider-like white noise levels, each with two frequency channels, assuming
 other frequencies are used for subtracting foregrounds. We
 have also made the simplifying assumption that in each experiment, the FWHM
 of both channels is the same as the channel with the larger beam. This does not affect
 the results much due to the crude size of the pixelization and the
 absence of a gravitational wave signal at small scales.  See Table~\ref{experiments} for other experimental assumptions.

For the Spider-like case we keep the flight time constant so that the observation gets deeper as $f_{\rm{sky}}$
decreases, while for the
Planck-like experiment the pixel noise is assumed constant for
different values of $f_{\rm{sky}}$.
The latter case, with small values of $f_{\rm{sky}}$, is used to illustrate how well a strategy of
only analyzing the lowest foreground sky could work, if for example, foreground removal turns out
to be prohibitive over much of the sky.  If foregrounds can be well-removed from Planck, then
 full sky is appropriate.

 We calculate the constraints on targeted cosmological parameters for different
 $f_{\rm{sky}}$'s, assuming  the observed patches are spherical caps from
 $\theta=0$ to $\theta=\theta_{\rm patch}$, corresponding to $\theta= {\rm
   cos} ^{-1}(1-2f_{\rm{sky}})$. 
  We perform the analysis  at different 
 resolutions for different sky cuts to minimize the effect of
 pixelization for small $f_{\rm{sky}}$ on the one hand,  and to keep the
 computational time reasonable for large $f_{\rm{sky}}$ on the other
 hand. We use $N_{\rm side}=32$, $N_{\rm side}=64$ and $N_{\rm side}=128$  for
 $f_{\rm{sky}}> 0.25$, $0.007<f_{\rm{sky}} \leq 0.25$,  and $f_{\rm{sky}}\leq 0.007$, 
 respectively. We checked the results for two neighbour resolutions at
 resolution switches. For the low $f_{\rm{sky}}$ switch, results are not
 sensitive to the change of resolution while for the switch at larger
 $f_{\rm{sky}}$  we are about $10\%-15\%$ pessimistic in the results by
 choosing the lower resolution, specifically for a Planck-like case (with small beam) and for a higher value of $r$, e.g., $r=0.12$. In these cases, lensing starts to dominate at higher multipoles and choosing a high enough resolution for the analysis would improve the errors on $r$ by resolving the primordial gravity waves at relatively high multipoles.
 
%-------------------------------------------------------

\subsection{Residual Foreground-Subtraction "Noise"} \label{sec:fgnd}

No study of gravitational wave detectability by $B$-mode experiments
can ignore the impact of polarized foreground emission. Component
separation is a major industry in itself. Various techniques have been
utilized with CMB data up to now - often involving template parameter
marginalization of one sort or another. We have been lucky so far in
that the foregrounds have been manageable for $TT, TE$ and $EE$. The
level of subtraction needed to unearth the very tiny gravity wave
induced $B$-signal is rather daunting, especially since the
foregrounds are largest at the low $\ell$. Thus, although we may
wrestle  the generalized noise from the detectors and from
experimental systematics to levels allowing small $r$ to be
detectable, the foregrounds will need to be well addressed before any
claim of primordial detection will be believable. Although we have
learned much already about the $TT$ foregrounds and, from WMAP, the
synchrotron $EE$, we do not know the $\ell$-shape or the amplitude of
the polarization for dust. In \cite{ode11,ode11_2}, the polarization emission
from thermal dust is based on a three-dimensional model of dust
density and  two-component Galactic magnetic field. It is assumed that
the degree of polarization has a quadratic dependence on the magnetic
filed strength and its direction is perpendicular to the component of
the local magnetic field in the plane of the sky, similar to the model
assumed by WMAP in \cite{pag07}. 
 In forecasting for proposed post-Planck satellite experiments, simple
 approximations for thermal dust and synchrotron emission have been
 made (e.g.,  \cite{bau09}, and references therein).  The dusty $\ell$-structure in this model is similar to the \cite{ode11} form:  ${\cal C}_{X\ell} \sim \ell^{-0.5}$ for $X=EE,BB$. We follow this \cite{bau09} approach here, but apply it to our pixel-based analysis.  

We therefore assume that the maps are already foreground-subtracted,
possibly with the wider Planck frequency coverage used in conjunction
with the Spider maps, with the CMB-component having a residual
uncertainty, which we incorporate in our analysis as an additional
large-scale (inhomogeneous) noise component $C_{{\rm N}}^{({\rm fg})}$. We  assume the power spectrum of the foreground residuals has the same shape as the original foreground spectrum, but with only  a few percent of the amplitude: 
\begin{eqnarray}
&& {\cal C}_{X\ell}\rightarrow {\cal C}_{X\ell}+\sum_{{\rm fg=S,D}} \epsilon_X^{({\rm fg})} {\cal C}_{X\ell}^{({\rm fg})}, \ X = EE, BB ,  \nonumber
\end{eqnarray}
with the sum over synchrotron S  and dust D emissions. The tunable removal-efficiency parameters  $\epsilon^{({\rm fg})}$ are taken  to be $5\%$ in our plots. 
The shapes are: 
\begin{eqnarray}
&& {\rm synchrotron:}\ {\cal C}_{X\ell}^{({\rm S})}(\nu)=\frac{\ell(\ell +1)}{2\pi} A_S \left( \frac{\nu}{\nu_0} \right)^{2\alpha_S} \left(\frac{\ell}{\ell_0} \right)^{\beta_S} \nonumber \cr
&& {\rm dust:}\ {\cal C}_{X\ell}^{({\rm D})}(\nu) = \frac{\ell(\ell +1)}{2\pi} p^2 A_D \left( \frac{\nu}{\nu_0} \right)^{2\alpha_D}  \left(\frac{\ell}{\ell_0}\right)^{\beta_D} \cr
&& \times \left[\frac{e^{h\nu_0/kT}-1}{e^{h\nu/kT}-1} \right]^2 \, .\nonumber
\end{eqnarray}
The dust polarization fraction, $p$, is assumed to
be around $5\%$. The values for the other parameters taken from \cite{bau09} are listed in
Table~\ref{fgparams}. They were  chosen to give agreement with WMAP, DASI and IRAS observations
(and the Planck sky model, which is based on these).  Although this model provides only a rough guide to the impact incomplete foreground subtraction will have on $r$-estimation, it does include the crucial large-scale dependence which differentiates it so much from the structure of the instrumental noise. 

%-------------------------
\begin{deluxetable}{ccc}
\tablecaption{Parameters of our assumed foreground model, adopted from \cite{bau09}.}
\tablehead{\colhead{Parameters} & \colhead{Synchrotron}&\colhead{Dust}}
\startdata
$A_{S,D}(\mu K^2)$ & $4.7\times 10^{-5}$& $1$ \\
$\nu_0$ & $30$& $94$ \\ 
$\ell_0$ & $350$& $10$ \\ 
$\alpha$ & $-3$& $2.2$ \\ 
$\beta^E$ & $-2.6$& $-2.5$\\
$\beta^B$ & $-2.6$& $-2.5$
\enddata
\label{fgparams}
\end{deluxetable}
%-------------------------------

A natural question when considering deep small sky observations is
how many patches there are on the sky with low foregrounds so the requisite cleaning is at a minimum.
The Planck Sky Model for the polarized foreground
emission \citep{lea08,del11} is similar to the one we have adopted. Using a code developed
by Miville-Desch\^enes, we have calculated for patches of radius $R$ the
pixel-averaged variance at pixel $p$, $\sigma_{\rm pol,fg}^2 (p, R)=
\avrg{(P -{\bar P (<R)})^2}$  of the
polarization intensity $P=\sqrt{Q^2 +U^2}$ about the patch-average
${\bar P}$ arising from the synchrotron and dust foregrounds. We
compare this with the $\sigma_{\rm pol,gw}^2(p, R)$ we obtained for each patch in a single
tensor-only primordial polarization realization (which is proportional to
$r^2$). The patches are sorted in decreasing order of the
"signal-to-noise" ratio $\sigma_{\rm pol,gw}(p, R)/\sigma_{\rm pol,fg}(p, R)$. The next pixel on the list
is included in a patch list if it has no overlap with the patches in the previously-determined higher signal-to-noise list.
A patch is considered to be $r$-clean if this polarization signal-to-noise exceeds unity, a rather strong criterion.
At $100 $ GHz, we found no "$r$=$0.01$"-clean patches, seven "$r$=$0.05$"-clean patches and ten "$r$=$0.1$"-clean patches with $f_{\rm{sky}} \gtrsim 0.007$ ($R=10^\circ$). There are one "$r$=$0.05$"-clean patch and two "$r$=$0.1$"-clean patches for  $f_{\rm{sky}} \gtrsim 0.03$ ($R=20^\circ$).
 At $150 $ GHz, we found no "$r$=$0.05$"-clean patches  and one
"$r$=$0.1$"-clean patch with $f_{\rm{sky}} \gtrsim  0.007$ but no $r$=$0.1$-clean patches for
$f_{\rm{sky}} \gtrsim 0.03$.

The non-overlapping criterion is quite severe. Another measure of
$r$-cleanliness is to determine the fraction of sky with
$\sigma_{\rm pol,gw}(p, R)/\sigma_{\rm pol,fg}(p, R)$ above unity. 
The "$r$"-clean fraction is clearly $\sim 0$ for those values of $r$ and $R$ with no corresponding clean patches (as stated above). Here only the non-zero values are reported.
At $100$ GHz, the "$r$=$0.05$"-clean fraction is $\sim 0.14$ ($R=10^\circ$) and the "$r$=$0.1$"-clean fraction is  $\sim  0.24$ ($R=10^\circ$); For both values of $r$, there is no appreciable decrease in the sky fraction by increasing the patch sizes to  $R=20^\circ$.
 At $150$ GHz, the "$r$=$0.1$"-clean fraction is  $\sim  0.04$ ($R=10^\circ$).
 It should be noted that as these sky fractions do not necessarily correspond to contiguous regions, the sky fraction of interest for small-sky B-mode experiments is in principle smaller.
 The Planck Sky Model at the lower frequencies agrees with the (extrapolated) synchrotron
emission from WMAP, but the higher frequency polarized dust emission really requires better observations, and awaits the release of the Planck mission results.
%-----------------------------------------

 \subsection{Correlations of $r$ with Other Cosmic Parameters}\label{sec:rcorr}
 
 Either detecting $r$ or placing a tight upper bound is crucial for progress in inflation studies. 
 Correlations of $r$ with other parameters $q^\alpha$ must be properly accounted for, since they are marginalized in the reduction to the 1D $r$-posterior. The relative importance of the various $q^\alpha$ 
 is determined by calculating the posterior-averaged cross-correlations $\avrg{\delta r \delta q^\alpha}_f $, which depend upon the experimental configuration and its noise. Within the quadratic approximation for the posterior information action, the correlations can be estimated from the inverse components, $[F^{-1}]^{r,\alpha}$, using the Fisher matrix equation(\ref{eq:Fab}), with lensing as well as instrumental noise included in the generalized noise matrix. Small steps in the main parameters of the standard
$\Lambda$CDM model $({\rm ln} (\Omega_{\rm b} h^2),{\rm ln} (\Omega_{\rm c} h^2) ,
H_0, n_{\rm s}, \tau, r)$ from the fiducial WMAP7 values\footnote{http://lambda.gsfc.nasa.gov/product/map/dr4/params/\\
lcdm\_sz\_lens\_wmap7.cfm} were taken to determine $F$ by numerical differentiation. 
The scalar amplitude $A_{\rm s}$ is treated as a normalization parameter here, so it is not included in the parameter list.
   We use two different fiducial values
for $r$,  $0.2$ and $0.01$, and three values of $f_{\rm{sky}}$, $0.007$, $0.07$ and $0.75$, for a Spider-like
experiment. 
We use a Gaussian prior on all parameters $q^\alpha$ but $r$
with  ${\bar q}^\alpha_{\rm i}$ and $w_{q{\rm i}}^{-1}= F_{\rm prior}$ given by the WMAP7  best-fit parameters. We choose $(F_{\rm prior})_{\alpha \beta} = \sigma_{\alpha,{\rm WMAP7}}^{-2} \delta_{\alpha \beta}$, which gives a weaker prior than the true WMAP7 results would give. In the quadratic approximation to the posterior information action, the total Fisher matrix is $F_t=F+F_{\rm prior}$. 

The average deviation in $r$, $\avrg{\delta r\vert \delta q^\alpha }$, and its  variance, $\avrg{\Delta \delta r \Delta \delta r \vert \delta q^\alpha }$, driven by given fluctuations in the other parameters, $\delta q^\alpha$, are
\begin{eqnarray*}
&&\bar{\delta r}\equiv \avrg{\delta r\vert \delta q^\alpha } =
  [F_t^{-1}]^{r,\alpha} [F_t]_{\alpha \beta} \delta
  q^\beta \\ \nonumber 
&&\avrg{(\delta r -\bar{\delta r})^2  \vert \delta q^\alpha } = [F_t^{-1}]^{rr} - [F_t^{-1}]^{r,\alpha} [F_t]_{\alpha \beta} [F_t^{-1}]^{r,\beta} \,\nonumber .
\end{eqnarray*}
If we let only one $\delta q^\alpha$ at a time differ from zero, and normalize the deviations to their 1-sigma values, we can express the result in terms of a dimensionless measure $ \rho_{r\alpha}$ of the degree of correlation:  
\begin{eqnarray*}
&&\rho_{r\alpha} \equiv [\avrg{\delta r\vert q^\alpha } /\sigma_r] /[  \delta q^\alpha /\sigma_\alpha ] \cr 
&&  \approx F_t^{-1}]^{r,\alpha} /\big([F_t^{-1}]^{rr} [F_t^{-1}]^{\alpha \alpha }\big)^{1/2}.
\end{eqnarray*}
The variance is $ \avrg{(\delta r -\bar{\delta r})^2  \vert \delta q^\alpha } \approx \sigma_r^2 (1-\rho_{r\alpha}^2)$. 
 
For the full sky case, we find the largest $\rho_{r\alpha}$ for $\tau$ and $n_{\rm s}$, with $\rho_{r \tau}$ and $\rho_{r n_{\rm s}}$  both $\approx 0.25$.
For  smaller sky coverage, the degeneracy between $r$ and
  $\tau$ disappears since the main constraints on $\tau$ come from the
  large scale polarization, which small cut-sky cases are not sensitive to. The dominant correlations of $r$ are with  the matter density parameters 
  $\Omega_{\rm c}h^2$ and $\Omega_{\rm b}h^2$, at the  $0.1-0.2$ level, a consequence of the  gravitational lensing induced $BB$ noise. 
  Even in the $25\%$ case for $\rho$, the constrained error diminishes only by $3\%$. 
 
Thus we should be able to safely estimate the error on $r$ with all or none of the basic six parameters
held fixed.  We verified this explicitly by comparing the 2D uncertainties calculated from the full $2$D $r-\tau$-grid 
with the full 6D uncertainties calculated from the inverse Fisher matrix, in $\ell$-space and in pixel-pixel space,  in Table~\ref{tab:uncertainty}, for different $f_{\rm{sky}}$ and at different resolutions, defined here by the value of $N_{\rm side}$. With all six parameters included,  $\sigma_r$ increases by only $\sim 10\%$ over the single  $\tau$-marginalized $\sigma_r$, which  justifies our exploration using a heavily  truncated parameter space to determine the errors in $r$. 

%----------------------------------
\begin{deluxetable}{ccccc}
\tablecaption{ $\sigma_r$ from the full likelihood computed on a 2D
  $r$-$\tau$ grid (bottom) cf. 1D, 2D and 6D Fisher determinations
  $[F^{-1}]^{rr}$ using pixel-space matrices (middle) and the
  simplified $\ell$-space sums, with $r_{\rm fid}=0.12$. This demonstrates that the use of reduced parameter spaces gives robust results, independent of cap sizes, here for $f_{\rm{sky}}=1,0.07,0.007$. }
\tablehead{ \colhead{method} & \colhead{param space}& \colhead{$N_{\rm
    side}=32$} &\colhead{$N_{\rm
    side}=64$} & \colhead{$N_{\rm
    side}=128$} \\
& & $f_{\rm sky}=1$ & $f_{\rm sky}=0.07$ & $f_{\rm sky}=0.007$}
\startdata
Fisher& 1 param & 0.022&0.018 &  0.037\\ 
$\ell$-space & 2 param & 0.023& 0.018& 0.037\\
& 6 param & 0.025& 0.020& 0.037\\ \hline
Fisher & 1 param & 0.022& 0.019& 0.034 \\ 
pixel-space& 2 param & 0.023& 0.019& 0.034 \\
& 6 param & 0.025& 0.020& 0.035\\ \hline
grid-based & 2 param & 0.021& 0.018& 0.036  
\enddata
\label{tab:uncertainty}
\end{deluxetable}
%-------------------------------
%--------------------------------------------------
\begin{figure}[h]
\begin{center}
\includegraphics[scale=0.52]{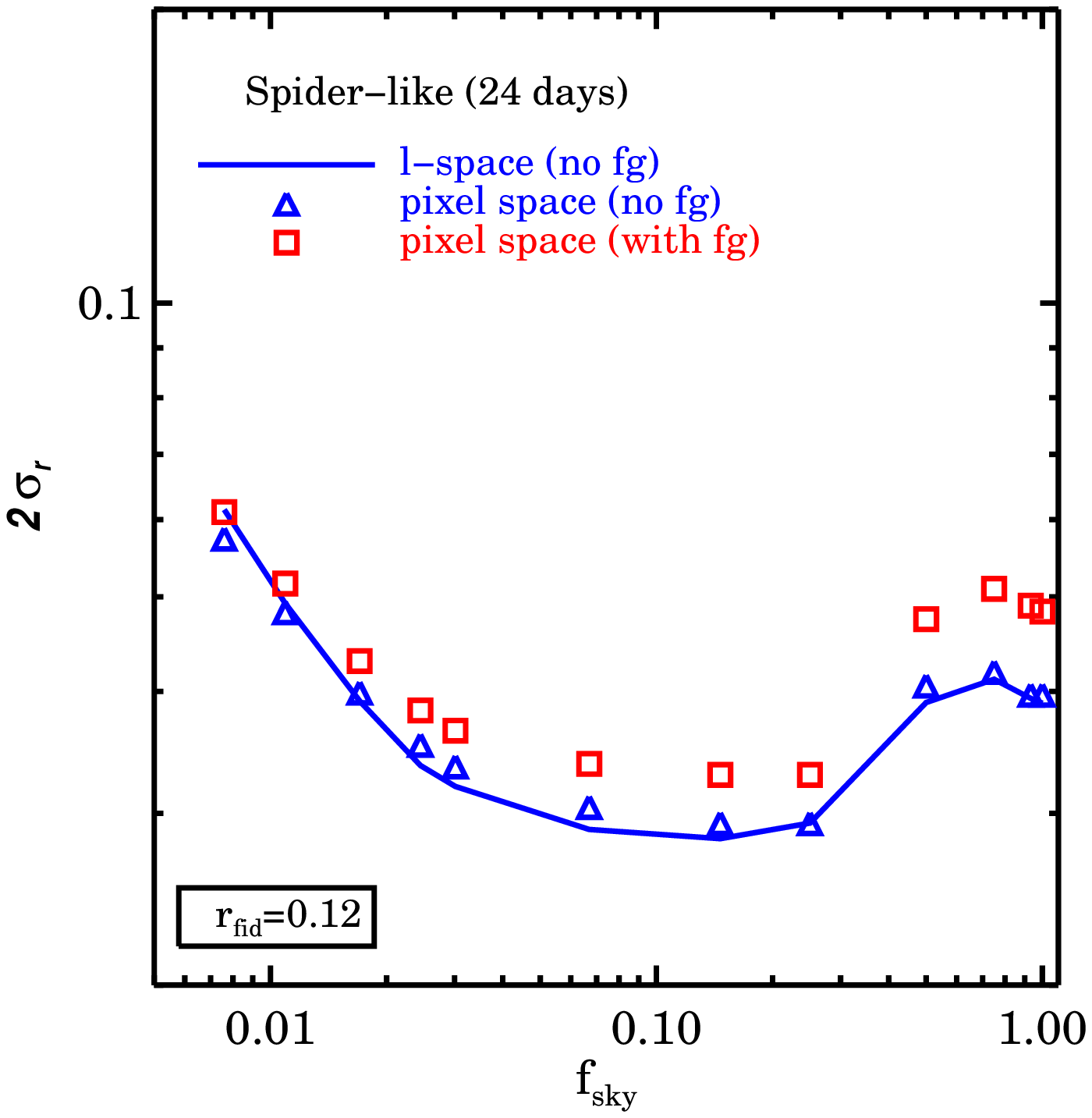}
\includegraphics[scale=0.52]{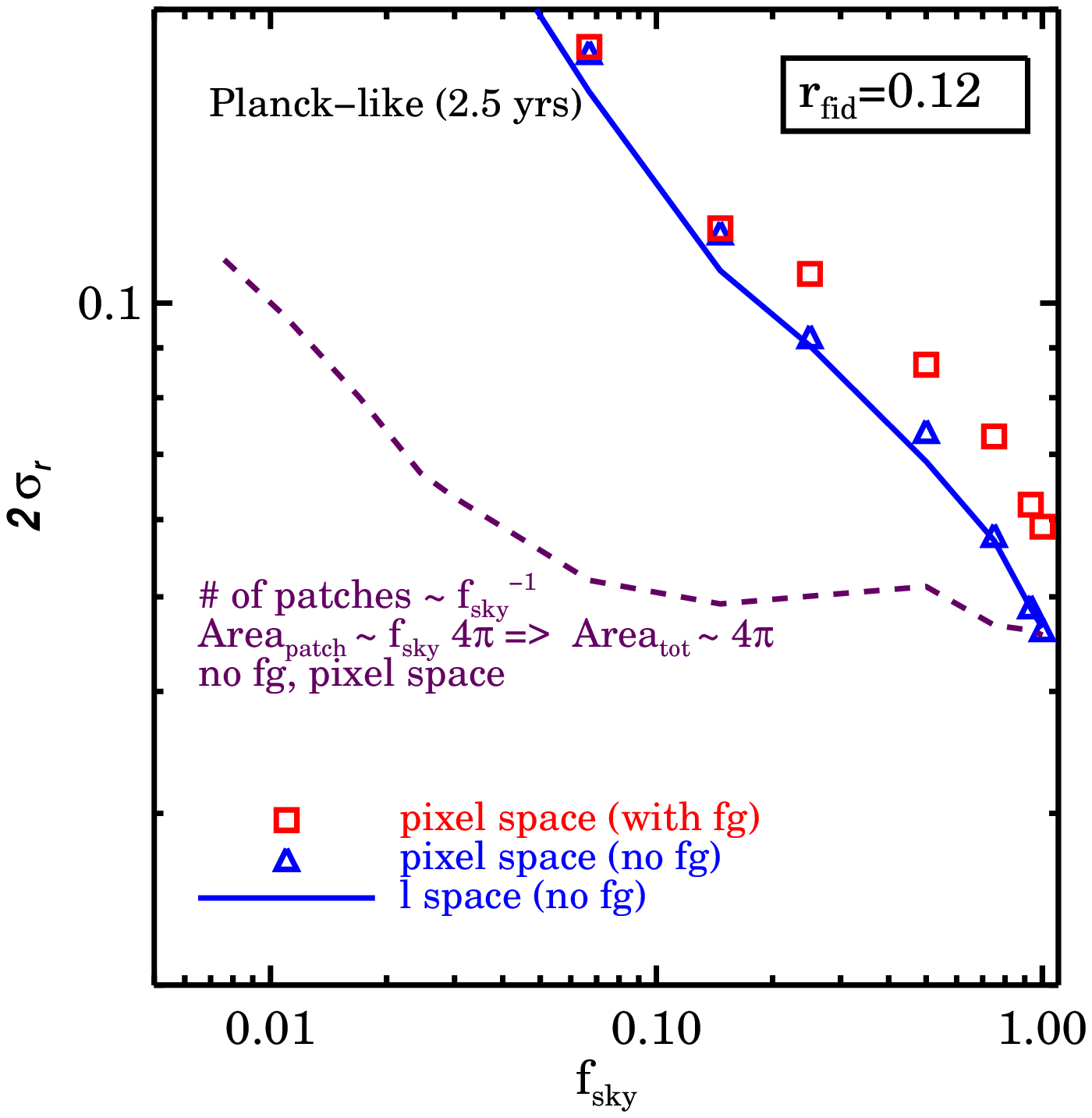}
\end{center}
\caption{Uncertainty in measuring $r$ for different sky
  coverages with Spider-like (top)  and  Planck-like (bottom) experiments, with and
  without foregrounds (squares and triangles respectively), for the fiducial
  model $r_{\rm fid}=0.12$. The solid lines are the results of  
  $\ell$-space analysis (ignoring
  foregrounds). The analysis has been performed with
  different resolutions for 
 different $f_{\rm{sky}}$, ranging from $N_{\rm side}=32$  for full sky to
 $N_{\rm side}=128$ for the smallest sky coverage. The $f_{\rm{sky}}$ refers to
 the sky coverage before applying the Galactic cut so for full sky
 $f_{\rm{sky}}$ is effectively $\sim 0.75$. The dashed line is the
 $2\sigma_r$ if the full sky needs to be effectively considered as a
 combination of several smaller patches with the individual
 observed sky fraction being $f_{\rm sky}$ and the total area of all
 patches  equal to Galaxy-masked  full sky.}
\label{SigmaBigr}
\end{figure}
%--------------------------------------------------

%--------------------------------------------------
\begin{figure}[h]
\begin{center}
\includegraphics[scale=0.52]{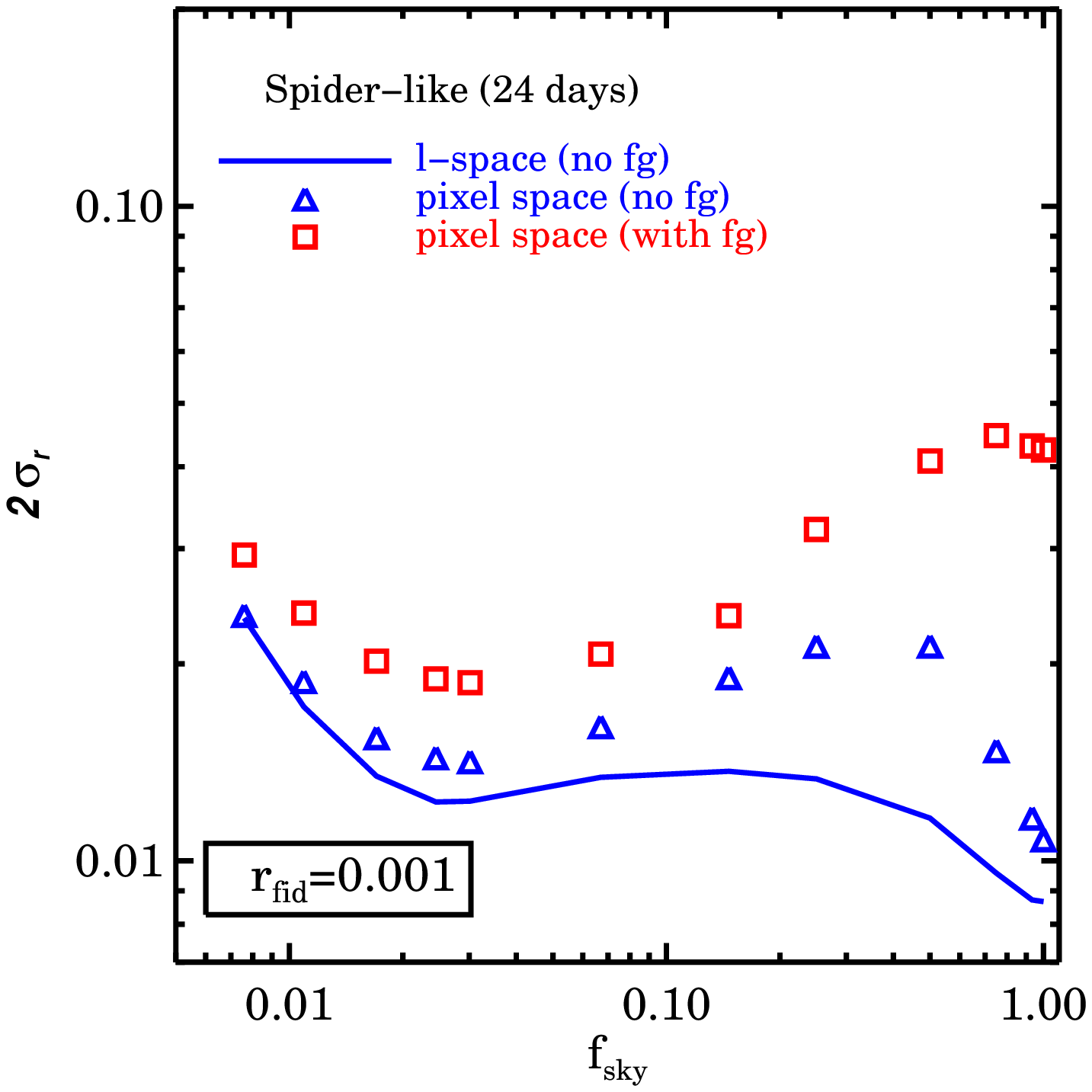}
\includegraphics[scale=0.52]{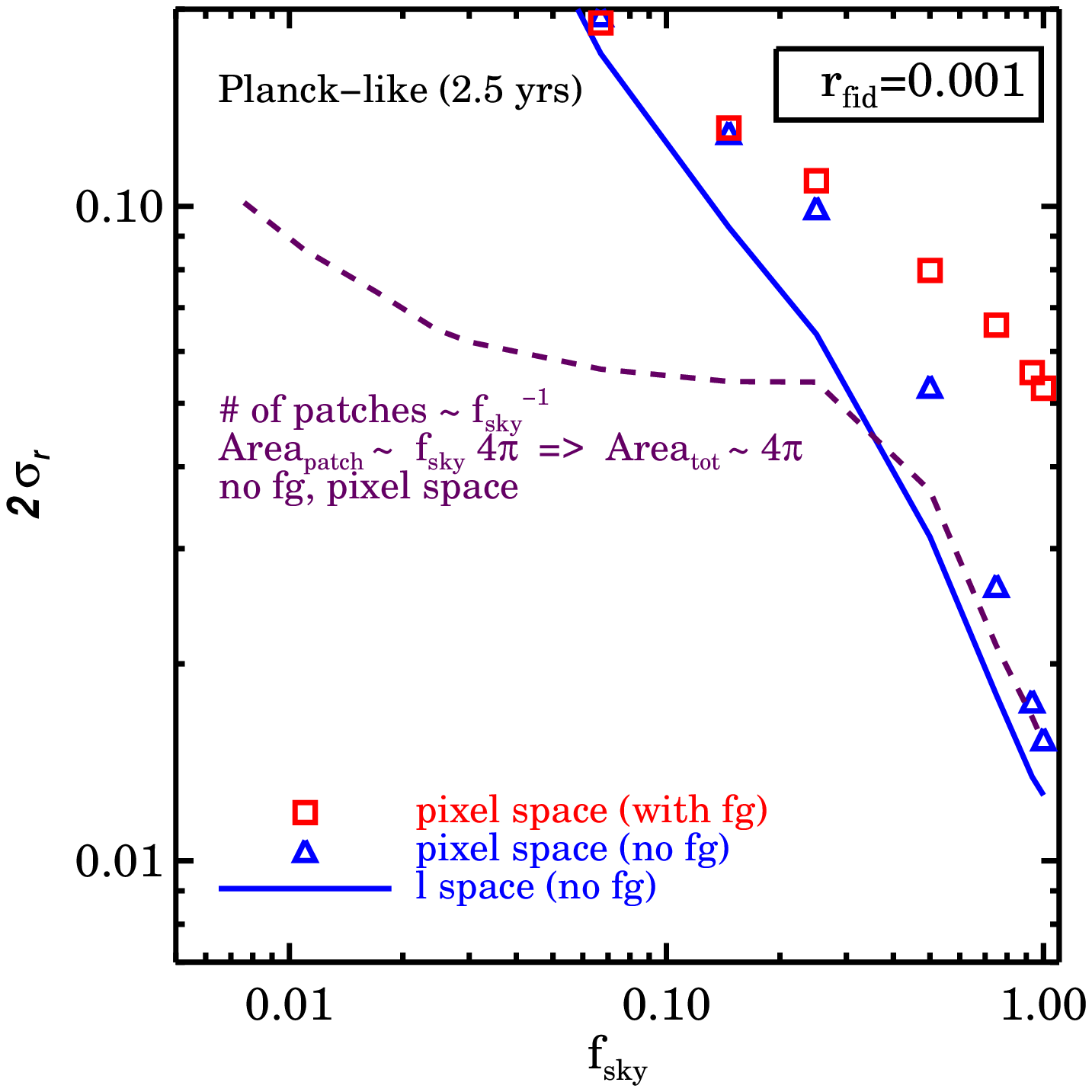}
\end{center}
\caption{Similar to figure \ref{SigmaBigr} with $r_{\rm fid}=0.001$.}
\label{SigmaLittler}
\end{figure}
%--------------------------------------------------

%----------------------------------------------------------------------
\subsection{Results in $r$--$\tau$ Space} \label{sec:rtau}
 
In this section, we use $\tau$ as well as $r$ to make our 2D parameter
 space since it has a direct impact on the $BB$ reionization bump.  
 We fix the overall ${\cal C}_\ell$ normalization for each parameter pair to the 
 WMAP $TT$ measurement at
 $\ell=220$. This is equivalent to having $A_{\rm s}$ as an adjustable
 parameter. If not otherwise stated, lensing has been included in all of the following
 simulations with a fixed noise template, linearly scaled with $A_{\rm s}$
 accordingly. 
Treating lensing in the noise covariance completely takes into account
its effect on sample variance. It may be possible for it to be partly
removed in the patch using delensing algorithms, (see
e.g., \cite{smi08} and references therein), leading to a reduced variance in the same way that we are treating a foreground residual.
However, treating
lensing as a noise source is a
 good assumption for our purposes here.

The $2\sigma_r (f_{\rm{sky}})$ plots in Figures \ref{SigmaBigr} and
\ref{SigmaLittler} are our main results. Shown are two  fiducial models with $r_{\rm fid}=0.12, 0.001$, both having $\tau_{\rm fid}=0.09$. The $f_{\rm{sky}}$ in the plots is 
the sky coverage before the Galaxy is masked. The Galaxy cut starts coming into the observed patch for
 $\theta_{\rm patch}\sim 40^\circ$. 
 
The results are compared to the expected error bars on $r$ from a simplified  $\ell$-space analysis. Counting modes properly is a difficulty in the $\ell$-space approximation for cut-skies. (This differs from  the full pixel-pixel covariance matrix analysis in which all modes are naturally taken care of.) 
For the $\ell$-space approximation, we have taken the mode number to be the naive $[f_{\rm{sky}}(2\ell+1)]$
where $[..]$ indicate the integer part. This imposes a low $\ell$-cut
on the modes by demanding $[f_{\rm{sky}}(2\ell+1)] \ge 1$ which overrides the $\ell$-cut from the fundamental mode of the patch, $2\ell +1 =[2\pi/2\sin(\theta_{\rm patch}/2)]$,  up to $\theta \approx 30^\circ $. 

This $\ell$-space $\sigma_r (f_{\rm{sky}})$ is a lower bound since it ignores the mode mixing on the cut sky.  Still, in the absence of systematic errors and 
for the simplified noise assumed here, the errors we find are near the true (matrix) values, as Figure~\ref{SigmaBigr} confirms  
for $r_{\rm fid}=0.12$. A similar measurement
with $r_{\rm fid}=0.2$ shows the same thing, though with a
more-flattened curve for $\sigma_r(f_{\rm{sky}})$ for the Spider-like case
and with foregrounds playing a smaller role. $E-B$ mixing does not seem to be a serious impediment, at least down to $f_{\rm{sky}}\approx 0.01$. 
For the Spider-like experiment, the error minimum is $2\sigma_r=0.035$ for $r_{\rm fid}=0.12$, at  $f_{\rm{sky}}\approx 0.15$, but the trough is broad. 
%--------------------------------------------------
\begin{figure}
\begin{center}
\includegraphics[scale=0.5]{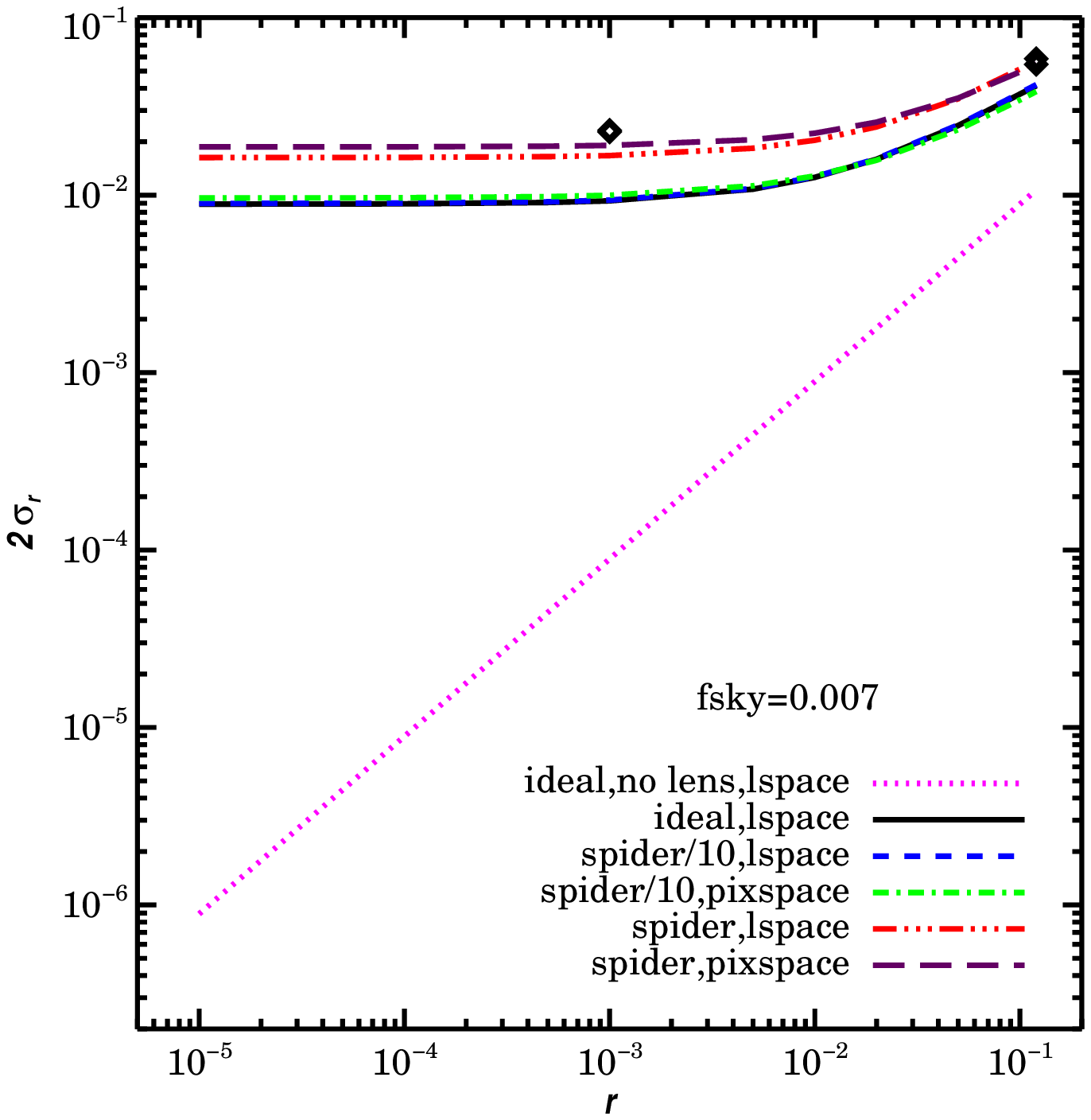}
\includegraphics[scale=0.5]{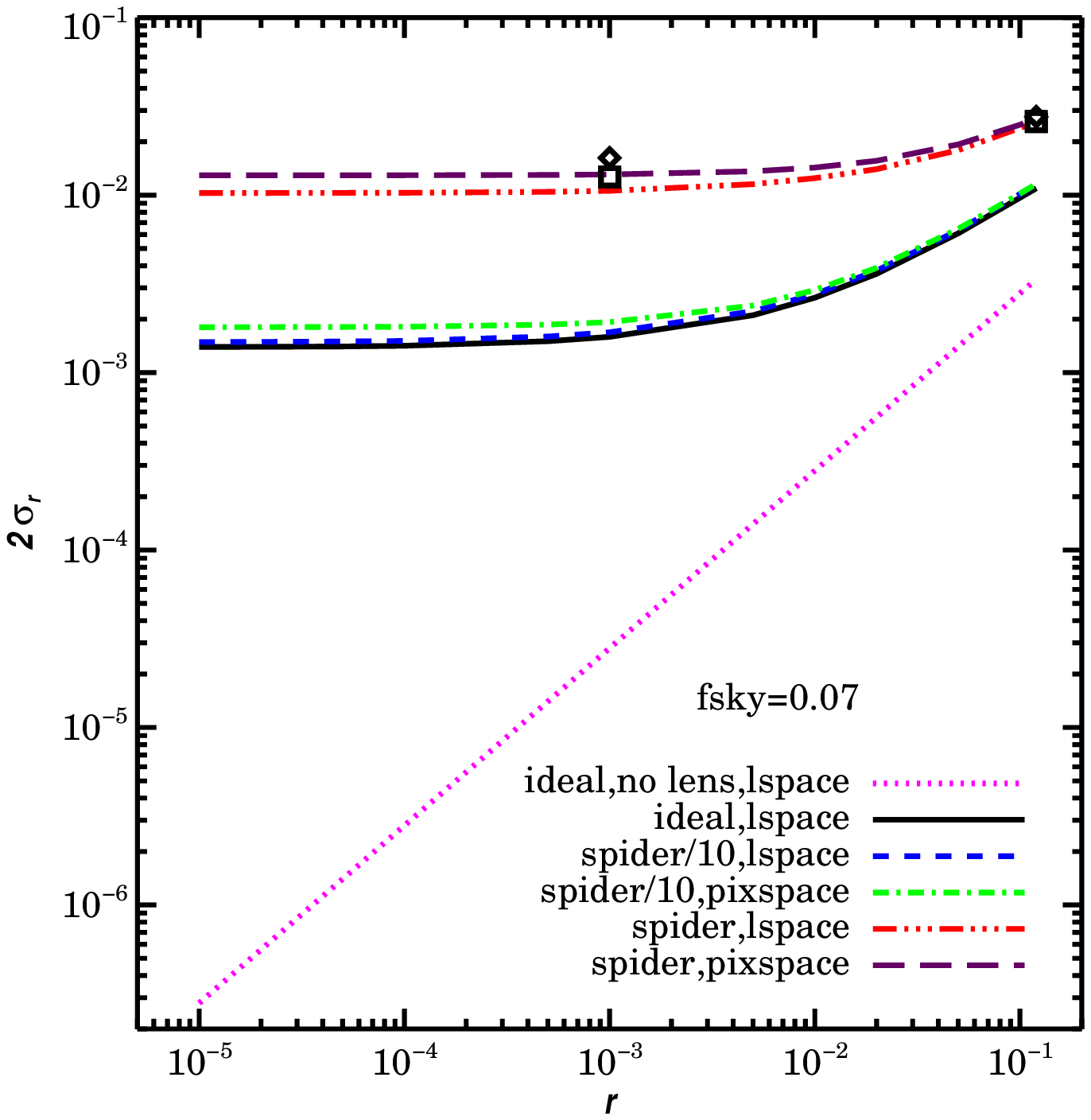}
\end{center}
\caption{The curves show $2\sigma_r$ as a function of $r_{\rm fid}$  obtained from the Fisher matrix in $\ell$ and
pixel-space for $f_{\rm{sky}}=0.007$ (top) and $0.07$ (bottom). The choices for the 
curves are meant to unravel the impact cosmic variance, lensing,
instrument noise and mode mixing have on $\sigma_r$. 
The symbols show errors from the full likelihood calculated on a gridded 2D parameter
space, and agree nicely for both pixel-space  (squares) and $\ell$-space (diamonds). }
\label{fish_sigma}
\end{figure}
%-------------------------------------------------- 
%----------------------------------------------------
\begin{figure*}
\begin{center}
\includegraphics[scale=0.7]{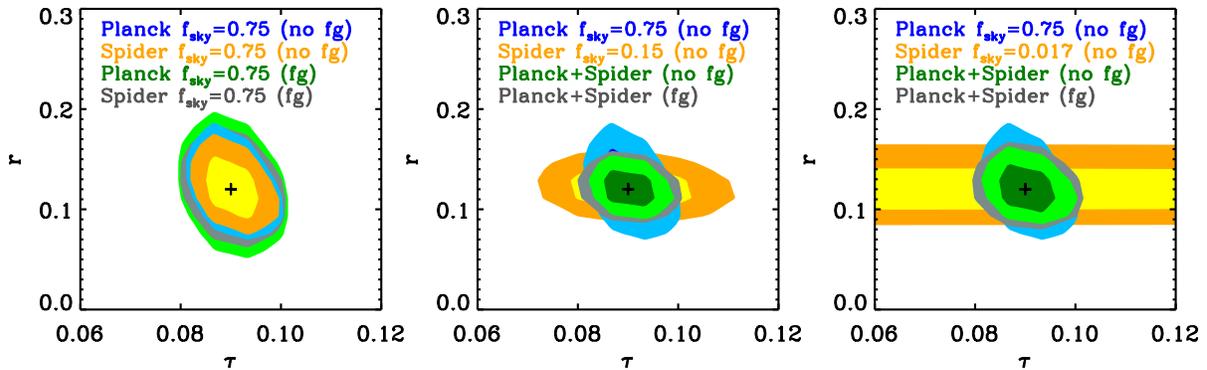}
\end{center}
\caption{$1\sigma$ and $2\sigma$ $r$--$\tau$ contours with and without
  foregrounds for a Spider-like experiment with different sky cuts
  and for a Planck-like Galaxy-masked experiment with effective
  $f_{\rm{sky}}\sim 0.75$. In the two right panels the contours for the combined
  Spider-like and Planck-like experiments are also plotted. The black plus signs
  denote the input $r_{\rm fid}=0.12$ and $\tau_{\rm fid}=0.09$. Expending Spider-like observing time on large sky coverage would not improve much the Planck forecasted  $\tau$ error, but would decrease the combined $r$ error, suggesting  the deep small-sky option is better.}
\label{contours_rtau}
\end{figure*}
%---------------------------------------
%---------------------------------------
\begin{figure}
\begin{center}
\includegraphics[scale=0.6]{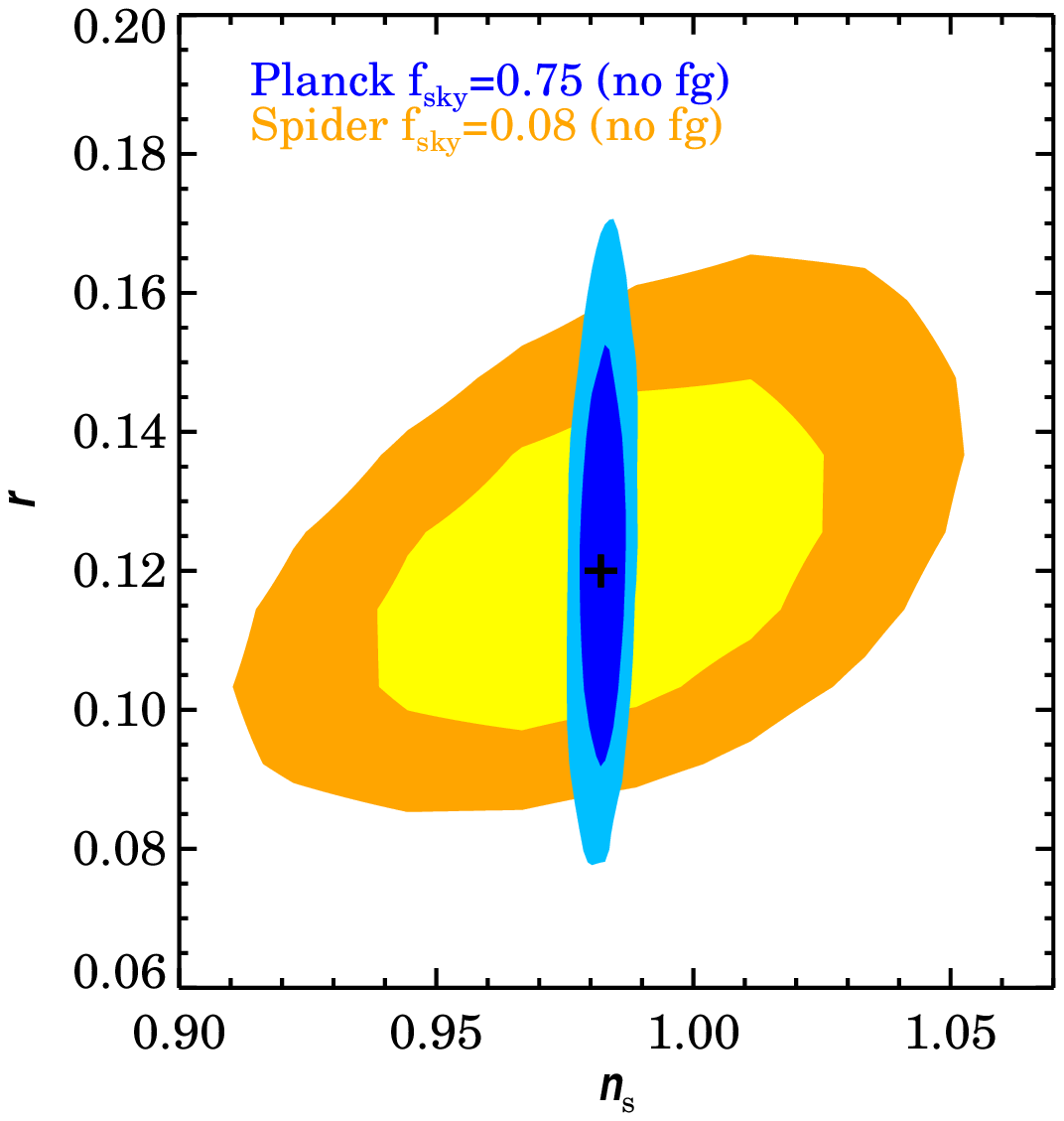}
\includegraphics[scale=0.6]{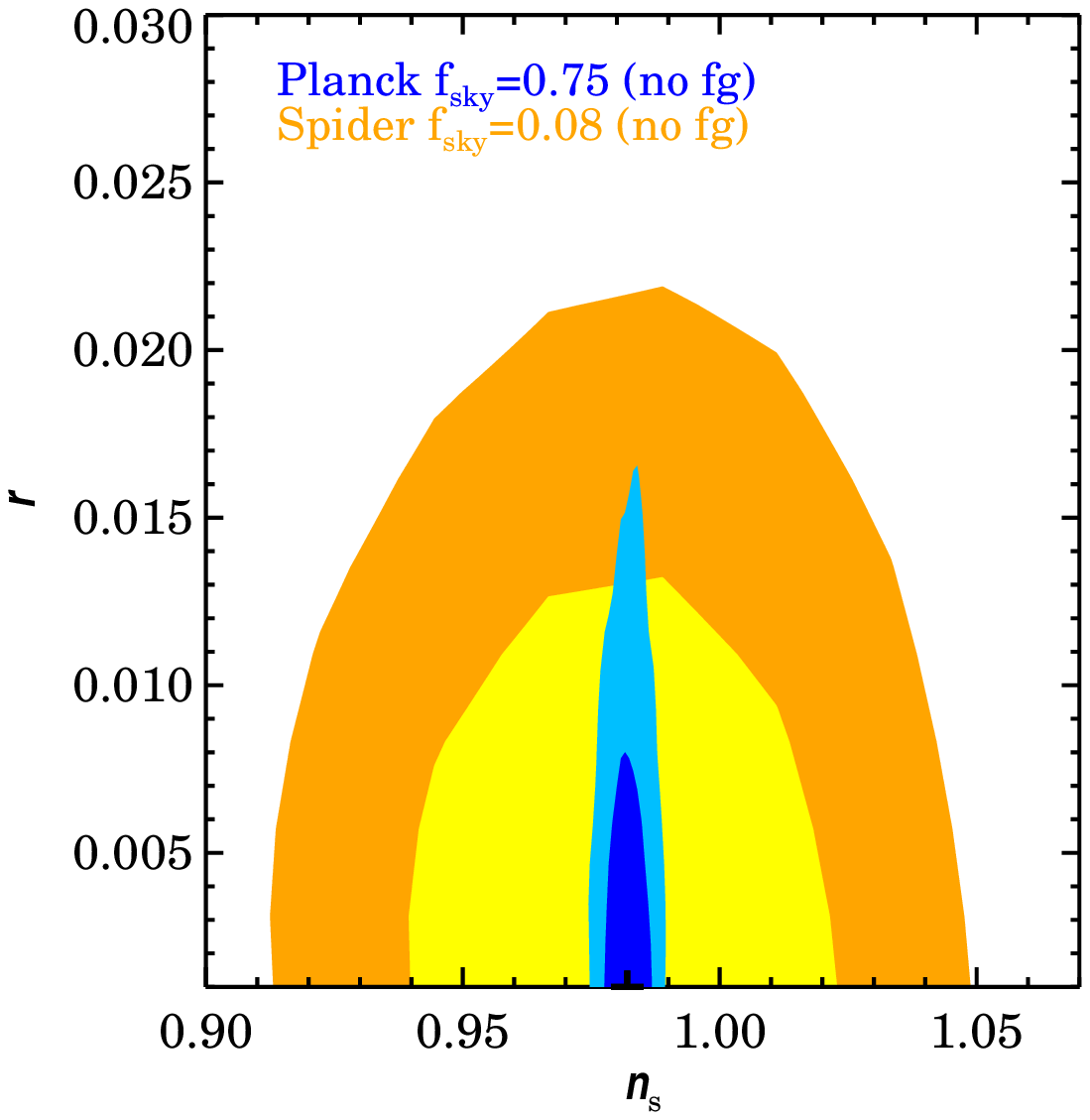}
\end{center}
\caption{$r$-$n_{\rm s}$ contours for a Spider-like
 $f_{\rm sky} =0.08$ experiment using pixel-space simulations are
  contrasted with that from a Planck-like Galaxy-masked $f_{\rm{sky}}=0.75$
  experiment. CosmoMC (http://cosmologist.info/cosmomc/) was used in the latter case to properly take into account the correlations of $n_{\rm s}$ with other cosmic parameters, which, unlike for $r$, are non-negligible.
  Top has $r_{\rm fid} = 0.12$ and bottom
  has 0.001; both have  $n_{{\rm s,fid}}=0.98$. Apart from
  demonstrating the small $\avrg{rn_{\rm s}}$, the plots indicate a
  possibly very rosy picture for constraining these two critical
  inflation parameters.}
\label{rns}
\end{figure}
%-------------------------------------------
\begin{figure*}
\begin{center}
\includegraphics[scale=0.7]{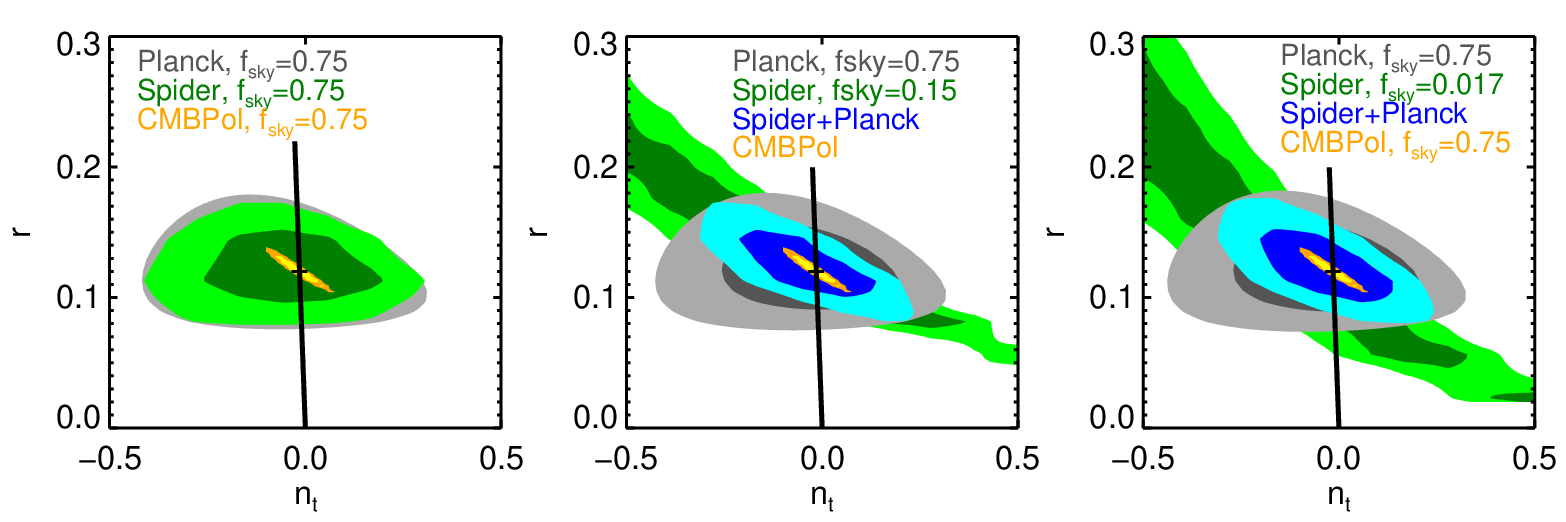}
\end{center}
\caption{$1\sigma$ and $2\sigma$ $r$-$n_{\rm t}$ contours for a Spider-like experiment with different sky cuts
  and for a Planck-like Galaxy-masked  $f_{\rm{sky}}$=0.75 experiment. The contours for a
  CMBPol-like experiment as well as those for the combined Planck-like
  and Spider-like experiments are plotted for comparison. The black
  line is the inflation consistency line and the black plus sign
is the fiducial input, $r=0.12$ and $n_{\rm t}=-0.015$. Even with this CMBpol, inflation consistency is not that well tested. }
\label{contours_rnt}
\end{figure*}
%--------------------------------------------------
%-----------------------------------------------------------
For the low $r_{\rm fid}=0.001$, for which only an upper limit can be expected, Figure~\ref{SigmaLittler} shows the agreement
in $\sigma_r (f_{\rm{sky}})$ between $\ell$-space and pixel-space is not quite as good, especially for $f_{\rm{sky}}\approx 0.25-0.5$ for which considerable observation time is expended on the $\ell\approx 12$ $BB$ valley
 (see Figure~\ref{window}) where
there is little signal. The naive $\ell$-space approximation
underestimates this, but  agreement with pixel-space is regained in runs with the reionization bump removed, by  setting $\tau=0$; 
for this case the monotonic rise in $\sigma_r (f_{\rm{sky}})$ with increasing $f_{\rm{sky}}$ continues to full sky.

Extending to the full Galaxy-masked sky improves the upper
limit on $r$  since the window function captures the
low-$\ell$ bump.  The $\ell$-space and pixel-space calculations disagree slightly, but when the  
Galaxy mask is removed, the estimates agree.

At small $f_{\rm{sky}}$, $2\sigma_r$ increases due to lensing which
dominates the total $BB$ spectrum at small scales.  The competition between
avoiding contamination by lensing and avoiding the $\ell\approx 12$
valley produces a weak minimum in $\sigma_r$ at $f_{\rm{sky}}\approx 0.15$
for $r=0.12$, when a detection is expected, and  at $f_{\rm{sky}}\approx 0.03$
for $r=0.001$, when an upper limit is expected.  The full sky is
weakly optimal for setting an upper limit in the absence of foregrounds.

The Planck-like measurements 
in the lower plots of Figures ~\ref{SigmaBigr} and ~\ref{SigmaLittler}
show a rise in $2\sigma_r$ as  $f_{\rm{sky}}$ drops wince the
information on the large scales are lost while the pixel noise stays
unchanged. The dashed lines in these plots show the approximate
$2\sigma_r$ for a full-sky Galaxy-masked Planck-like experiment if the large-scale modes are filtered e.g., by time-domain
filtering or due to high foreground contamination and thus the
observed region is considered to be a combination of smaller patches
(adding up to the full sky in total observed area).

Not surprisingly, we see that foregrounds mostly affect experiments
with larger $f_{\rm{sky}}$, and for fiducial models with smaller $r$.  We also see that deep observations of quite
small patches seem to do as well as larger patches (observed less
deeply) and even much better if $r$ is small (for which  the sample variance is
very small and instrument noise plays the dominant role).

Figure~\ref{fish_sigma} shows how different components contribute to
the error on $r$ calculated using the Fisher matrix for various $r_{\rm fid}$ and $f_{\rm{sky}} = $ $0.007$ and $0.07$. 
As before the mode mixing is ignored in the $\ell$-space
calculation. If there were no
lensing and no mode-mixing, in the limit of no instrument noise, the only source of error would be
the sample variance, which is,  as expected, proportional to $r$. The
solid black lines show the minimum irreducible errors due to sample
variance and lensing. We contrast this with calculations in both pixel and
$\ell$-space of two Spider-like experiments. One has 10 times less noise than the fiducial Spider case, a noise level that  
can be seen to give almost no contribution to the errors for these sky cuts since lensing noise is dominant. The other has our standard Spider-like noise, which can be seen to  
 significantly add
to the error. The neglect of mode-mixing in determining $\sigma_r$ vanishes as $r$
 increases, since sample variance dominates the error, as a comparison of the curves from the pixel-space and $\ell$-space
analyses shows.  The over-plotted
 symbols represent the errors from measuring the likelihood curve in a
 gridded 2D parameter space (as explained earlier). The $2\sigma_r$'s from the full method and the Fisher matrix approximation are close. The small difference is because the
  $r$-likelihood curve is not a perfect Gaussian.

Figure~\ref{contours_rtau}  shows the $2$D $r$--$\tau$ contours for $3$ different
values of sky coverage for a Spider-like experiment compared 
to a full-sky Planck-like experiment (with Galaxy mask
cut) with and without foreground contamination.  
 As expected, $\tau$ is unconstrained as $f_{\rm{sky}}$ is decreases for the Spider-like experiment since $\tau$-constraints come from the largest
angular scales: what is optimal for $r$ detection is awful for  $\tau$ determination, for which all-sky is best. 

%---------------------------------
\subsection{Results in $r$--$n_{\rm s}$ Space} \label{sec:rns}

In Figure~\ref{rns},  we have plotted the $r$--$n_{\rm s}$ contours for an $f_{\rm{sky}}=0.08$ 
Spider-like experiment and for a full-sky Planck-like
survey, with and without foregrounds, using the model discussed in \S~\ref{sec:rtau}. This shows almost no correlation between the
two parameters for these experimental cases, as expected from the
discussion in \S~\ref{sec:rcorr}. It also shows the remarkable set of inflation constraints that may arise from Planck and Spider-like experiments. 

%------------------------------------------------------------
  \subsection{Results in $r$--$n_{\rm t}$ Space} \label{sec:rnt}
  
 Although detecting $r$ would provide an invaluable measure of the mean acceleration parameter (and energy scale) of
  inflation, we want more, the shape of the tensor power embodied in
  the tensor tilt $n_{\rm t}$, which we explore here in a 2D space 
  by fixing $\tau , n_{\rm s}$ and the other cosmic parameters.  Figure~\ref{contours_rnt}
  shows the 2D contours for $r$--$n_{\rm t}$ with $r_{\rm fid}=0.12$,
  and fiducial tensor tilt $n_{{\rm t,fid}}=-0.0150$ satisfying the
  inflation consistency condition eq.~\ref{eq:rnt}. Alas, we see that
  $n_{\rm t}$ is hardly
  constrained by  Spider-like and Planck-like experiments, no matter how large $f_{\rm{sky}}$ is.  To see whether a post-Planck deep all-sky 
  experiment could modify this conclusion comparison, we ran our
  analysis using the specification of a putative mid-cost CMBPol
  mission outlined in \cite{bau09}, using the frequency channels
  described in Table~\ref{experiments}. There is of course improvement, and the COrE and PIXIE post-Planck missions would do better, but the relatively short $\Delta \ell \sim 150$ baseline precludes even an ideal experiment from providing a  powerful test of inflation consistency. 

%-----------------------------------------------------

\subsection{Breaking $r$ up into $r_{X\beta}$-Shape Parameters: A Tensor Consistency Check }\label{sec:rXbeta}

Because $r$ is essentially a linear parameter (for given $A_{\rm s}$), we are effectively determining a single (very) broadband power amplitude multiplying a 
collection of fiducial $X$-template shapes ${\cal C}_{X\ell}^{(\rm g)}$ given by the gravitational wave powers. It is natural to test this locked-in monolithic parameterization by introducing a collection of parameters $r_{X\beta}$ multiplying individual $X$ and $\ell$-band templates: 
\begin{eqnarray}
&&{\cal C}_{EE\ell} ={\cal C}_{EE\ell}^{({\rm s})}+r_{EE\beta} \chi_\beta (\ell ) {\cal C}_{EE\ell}^{(\rm g)}\cr
&&{\cal C}_{BB\ell} ={\cal C}_{BB\ell}^{({\rm lens})}+r_{BB\beta}\chi_\beta (\ell ) {\cal C}_{BB\ell}^{(\rm g)}  \, .
\end{eqnarray} 
Here ${\cal C}_{EE\ell}^{({\rm s})}$ is the scalar part of ${\cal
  C}_{EE\ell}$, including lensing,  and ${\cal C}_{BB\ell}^{({\rm lens})}$ is the lensed BB power.  The overall normalization is arranged so that $r_{X\beta}=r$ is the tensor consistency condition. 
The $\chi_\beta (\ell )$'s  are the $\beta$-windows. These have often been taken to be top-hats satisfying a saturation property $\sum_\beta \chi_\beta (\ell ) =1$ and an orthogonality property $ \chi_\beta (\ell )\chi_\beta^\prime (\ell )=\delta_{\beta \beta^{\prime}}$ in bandpower work. However, the modes could also be quite overlapping as long as saturation and the $r_{X\beta}=r$ normalization are satisfied. 

\begin{figure*}
\begin{center}
\includegraphics[scale=0.7]{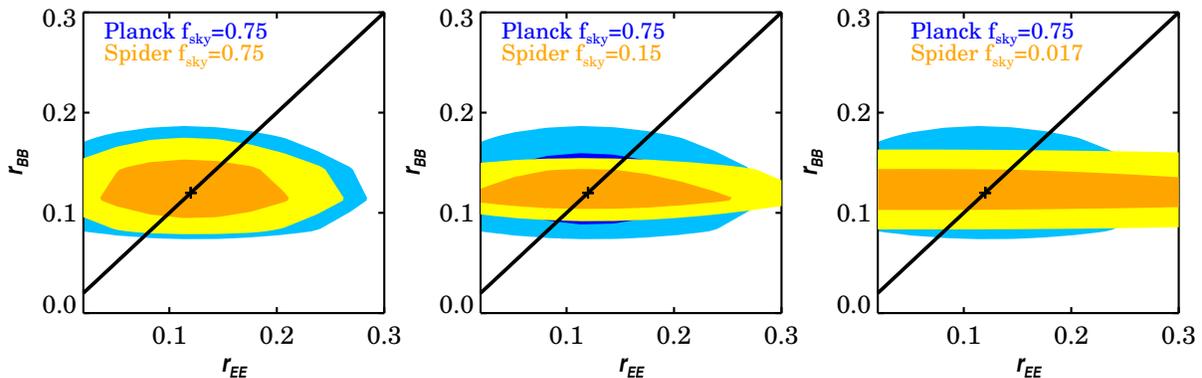}
\end{center}
\caption{$1\sigma$ and $2\sigma$ contours in the $r_{EE}$--$r_{BB}$ plane for a Spider-like experiment with different sky cuts
  and for a Planck-like experiment with $f_{\rm{sky}}$=0.75. The black solid lines show the {\em tensor consistency}  curves $r_{EE}=r_{BB}$  and the plus signs
  show the fiducial $r_{EE}=r_{BB}=0.12$ input model. As expected, $r_{BB}$ is better determined than $r_{EE}$ and this tensor consistency is not well tested. }
\label{contours_qeb}
\end{figure*}

This is a reasonable path to finding the tensor bandpowers for $BB$
and $EE$ but, given the \S~\ref{sec:rnt} result on $n_{\rm t}$, we will content ourselves with a 2D example using one $\ell$-band $\beta$ and two $X$ parameters,  $r_{EE} $ and $r_{BB} $.  For this study, we keep $A_{\rm s}$ fixed (cf. \S~\ref{sec:rtau} and \ref{sec:rnt}). The contours in Figure~\ref{contours_qeb} show the degree to which the tensor consistency encoded in the $r_{EE}=r_{BB} $  line, can be checked.  The contours confirm the expectation that the  $B$-modes are the most influential source of information about primordial tensor perturbations, since the large scalar contribution to $EE$ swamps the tiny tensor signal, inflating the error
bars. Using checks like these for showing consistency have had a long history. In the first $EE$ polarization detection papers, the EE amplitude was shown  to be consistent with the  amplitude expected from $TT$ parameters \citep{kov02,sie04}. In the first lensing detections in the TT power spectra, the deviations from lens-free results were shown ro be consistent with expectations from the parameters determined from the primary TT data \citep{rei09,dun10}.  

%--------------------------------------------------------
%----------------------------------------------------------

\subsection{Breaking $f_{\rm{sky}}$ into Many Fields}\label{sec:4patches}

\begin{figure}[h]
\begin{center}
\includegraphics[scale=0.55]{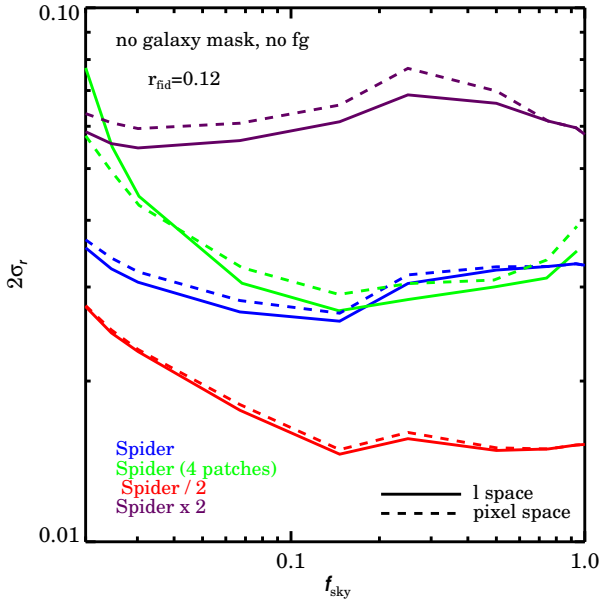}
\end{center}
\caption{When one patch covering $f_{\rm{sky}}$ is broken up into four $f_{\rm{sky}}/4$ cap-patches, but the noise and observing time remain constant, 
the ($\tau$-marginalized) $r$-errors remain similar except at very small $f_{\rm{sky}}$. We also show that factors of two changes in the noise swamp this effect. The calculations were done with $r_{\rm fid}=0.12$ in the pixel-space except for the
  highest sky coverages where the pixel and $\ell$-space analysis are
  in excellent agreement. The effect of foreground contamination and  Galaxy
  cut has not been taken into account here.}
\label{fig:4patch}
\end{figure}

Using multiple (foreground-minimized) fields to make up a total $ f_{\rm{sky}}$ is an approach that has been advocated for 
ground-based strategies (e.g., for ABS, \footnote{http://www.princeton.edu/physics/research/cosmology-experiment/abs-experiment/}). In Figure~\ref{fig:4patch} we show the impact of splitting $ f_{\rm{sky}}$ into four patches, while keeping the total integration time and the instrument noise constant. One does not lose that much as long as the total probe is a few percent of the sky, a consequence of the broad single-patch $\sigma_r ( f_{\rm{sky}} )$ minimum. 
The number of polarization-foreground-clean patches is of course still
to be determined. We also varied the patch geometry; e.g., for an
$f_{\rm{sky}} \sim 0.08$ rectangular region with $r_{\rm fid}=0.12$, we get $ 2\sigma_r=0.048$ without foregrounds, in good agreement with the cap result $2\sigma_r=0.050$. 

%------------------------------------------
%-------------------------------------------------

%-----------------------------------------------
\section{Summary and Conclusions}\label{sec:concl}

In this paper, we applied a full matrix likelihood analysis to multifrequency $Q$-$U$ polarization maps and $T$-maps of forecasted data to determine the posterior probability distribution of $r$. 

\subsection{Leakage Levels and Leakage Avoidance}\label{sec:avoidleak}

This method avoids the explicit linear $E$-$B$ decomposition of the
polarization maps before doing the likelihood analysis and gives the
best possible determination of $r$, provided that systematic errors
are correctly modelled.  For realistic cut-sky observations,  we
measured the level of $BB$ contamination from the inevitable
mode-mixing from the much larger $EE$  power. In addition, there is
leakage from instrumental effects, in particular with $T$ seeping into
$Q$ and $U$, which has to be included in any approach. We have left
the investigation of this issue to future work.

%-----------------------------------------------------
\subsection{Computational Feasibility of Exact Likelihoods}\label{sec:complike}

It is often the case in CMB cosmology that the shear number of pixels precludes a direct  full map-based likelihood procedure, with an intermediate power spectrum determination done before parameter estimation. However, for Spider and similar ground and balloon experiments targeting $r$, relatively low resolution and restricted sky coverage are all that is really needed for detection.  The  result is a total pixel number that allows computationally feasible inverse and determinant calculations of the large signal-plus-noise correlation matrices  $C_{t }= C_{\rm N} + C_{\rm S}(q)$ -- with contributions from both the parameter-dependant signal covariance $C_{\rm S}(q)$ and the generalized noise $C_{\rm N}$, which includes uncertainties from the foreground subtraction as well as from instrumental and systematic noise in the maps

Matrix methods have had a long history, dating from the earliest CMB data sets, e.g.,  \cite{bon01}. For example, they were used for COBE, Saskatoon, Boomerang, and CBI analyses. Often compression was used, e.g., to signal-to-noise eigenmodes \citep{bon95,bon01} or by coarse-grained gridding \citep{mye03}, to make the matrix manipulations tractable. With Boomerang, an important aspect was to make sure all issues regarding data-filtering, inhomogeneous and aspherical beams, transfer functions, striping etc. were properly included. Invariably, a Monte Carlo simulator of each experiment has been built, in which simulated timestreams have as many effects from systematic and data processing as one can think of included. 

%---------------------------------------------------
\subsection{Matrix Estimation from Monte Carlo Noise and Signal Simulations and Relation to Master/XFaster}\label{sec:matrixest}

The Master/XFaster approach encodes this in isotropized $\ell$ space filters and rotationally symmetrized masks which allow one to relate the underlying all-sky $C_{{\rm S},cX\ell}$ to the filtered cut sky. Similarly an isotropized noise $C_{{\rm N},cX\ell}$ is also determined by taking processed noise timestreams, creating maps with them, $Y_{\ell m}$ transforming them, then forming  a quadratic  average over noise samples $J_s$, $C_{{\rm N},cX\ell} = \sum_{J_s,m } \vert a_{NJ_s,cX\ell m}\vert^2/[(2\ell +1)N_{s}]$. 

When one has a large number of detectors, using only cross-correlations and no auto-correlations has an advantage, namely that the the cross-noise is small, from systematic effects in the arrays and instrument as a whole. Precise modelling of the auto-noise is not easy. However, any operation that can be done for Master or XFaster can also be done to estimate the noise matrices, using noise sample sums. (Getting convergence of small off-diagonal components may require many samples). Matrices have the advantage that they naturally allow for anisotropic and inhomogeneous components, in the noise maps - including striping effects -  and in the beam maps and in the foreground maps. There are issues about optimal estimation of  the generalized pixel-pixel matrices that one would like to tune, but there are no fundamental obstacles to making the $C_{\rm N}$ and $C_{\rm S}$ matrices highly accurate for parameter estimation. 

WMAP used a matrix-based likelihood for low $\ell$, connected to an isotropized $\ell$-space likelihood covering the high $\ell$'s. Planck is doing the same. We expect such a hybridized likelihood code will also be used for Spider-like experiments for routine parameter estimation, even though we think one can get away with a full matrix likelihood code. 

If simulated timestreams are used for $C_{\rm N}$ and $C_{\rm S}$
estimation, generalized pixels may prove preferable to the usual
spatial pixels. The Cosmic Background Imager  CBI \citep{mye03,sie04} used the reciprocal space pixels for the primary construction, rather natural for an interferometry experiment where the timestream analog is a set of visibilities. ACT and QUaD  also have done their power spectrum estimation in the Fourier transform space of spatial maps. 
%-------------------------------------------------------

\subsection{The CBIpol Approach as a Guide for Small Deep-sky Analyses}\label{sec:egCBIpol}

The use of  matrix likelihood codes does not mean that $E$ and $B$  maps will not be constructed, just that parameters would not be extracted from them. The CBI example of how such $E,B$ maps were made and used, and why bandpower and parameter estimations did not use $E,B$ maps serves as a paradigm for how things could proceed for Spider-like data. The CBI data were compressed (via a GRIDR code) onto  a discrete (reciprocal) lattice of wavenumbers by  projecting measured interferometer visibilities onto a gridded 2D ${\bf K}$-space. A direct unitary transformation takes such a basis of "momentum" modes into a basis of spatial modes in real space where $Q$-$U$ is a  more appropriate representation. An important point is that the polarization map estimators evaluated on the discrete wavenumbers of the lattice are linear combinations of the continuous wavenumbers, the mode-coupling of finite maps which also leads to an $E$-$B$ mixing.  

In the lattice representation, the resulting size of the correlation matrices for CBI were quite tractable for direct inversion and the full likelihood was evaluated (via an mLikely code) to determine bandpowers for $TT$, $EE$, $BB$ and $TE$, without separation of the Fourier maps into  $E$ and $B$. 

An optimal linear map reconstruction of  $E$ and  $B$ was done for visualization purposes, with real-space and momentum-space maps showing the CBI   $E$ and $B$  Wiener-filtered means, accompanied by a few maps showing typical fluctuation maps about the mean maps. These were contour maps, since the usual headless vector polarization plots are of length the polarization degree, $\sqrt{Q^2+U^2}$, tilted at an angle $\arctan(U/Q) /2$. 

For Spider-like bolometer-based experiments for which the raw data are bolometer time-streams from which $QU$ maps are constructed, the compression step leads to tractable matrices as in the CBIpol case, although in the first instance the pixelization choice may be in real space  rather than in wavenumber space or in a generalized-pixel space. Just as with CBIpol, parameters and bandpowers would be determined with direct likelihood calculations, yet Wiener-filtered $EB$ maps would still be made for visualization. 

%---------------------------------------------------------
\subsection{Exact $2$D Likelihood Computation}\label{sec:rtauresult}

Given the matrix construction method, we determined the  posterior probabilities on reduced $2$D-grids consisting of $r$ and one other cosmic parameter, in many cases  the Thomson scattering depth to reionization $\tau$. The grid could be extended to higher dimensions, as they were in early CMB analyses of COBE, Boomerang, CBI and ACBAR. More efficiently,  MCMC chains could be used to explore the posterior probability surface.  Since, as we have shown,  $r$ is relatively  weakly correlated with the other standard cosmic parameters, our use of a reduced dimensionality is accurate. We targeted $\tau$ for a second parameter, although it too is weakly correlated for Spider-like experiments probing modest $f_{\rm{sky}}$, because of its importance for the reionization bump in $BB$  which is picked by large $f_{\rm{sky}}$ experiments such as Planck. We showed that as long as the input value $r_{\rm fid}$ is reasonably larger than the error $\sigma_r$, e.g., $\sim 0.1$, $r_{\rm fid}$ can be well-recovered by our methods.  

%-----------------------------------------------
\subsection{The Inflation and Tensor Consistency Checks}\label{sec:inflconsist}

We have used $r$ and $n_{\rm t}$ for our reduced 2D parameter space to see
how well the inflation consistency condition, $n_{\rm t} \approx - r/8$, can
be tested. For example, with $r_{\rm fid}=0.12$ and the consistency
value $n_{\rm t,fid}=-0.015$ , we obtain $2\sigma_r  \approx 0.036$ and
$2\sigma_{n_t}  \approx 0.28$. The large 1-sigma error on $n_t$
is what one might have expected given the relatively small
$\ell$-baseline (reminiscent of the $\pm 0.2$ limit on $n_{\rm s}$ from the
even smaller baseline COBE DMR data). Thus, although breaking up $r$
into bands will be useful, the $n_{\rm t}$   slope that follows will
be not be powerful enough to test consistency. With CMBpol and at
$N_{\rm side}=512$, the errors are $2\sigma_r  \approx 0.014$ and
$2\sigma_{n_t}  \approx 0.07$ , still too large. A more prosaic internal consistency check was done to show that what one thinks is $r$ from the total $BB$ agrees with what one gets from the less-tensor-sensitive total  $EE$. 

%--------------------------------------------------------------
\subsection{Relation to Planck}\label{sec:cfPlanck}

We based our Planck-like case on the Blue Book detector specifications. The actual in-flight performance is quite similar \citep{hfi11,lfi11}. It is encouraging that five full sky surveys of six months seems possible, as we near the end of the fourth. What will emerge from the actual Planck polarization analysis may be quite different from the simplified foreground-free $2\sigma_r (f_{\rm{sky}} =0.75)\sim 0.015$  forecast of white experimental noise with well-subtracted foregrounds of known residual, and with no systematics. This relies on the $BB$ reionization bump being picked up, but the required low $\ell$'s  are especially susceptible to the foreground-subtraction residuals ($2\sigma_r (f_{\rm{sky}} =0.75)\sim 0.05$) and systematic effects. Some of the issues are described in \cite{efs09}. Irrespective of how well Planck wrestles with the low $\ell$ issues, it will be able to analyze many patches within the $ 75\%$ of the sky, rank-ordered by degree of foreground contamination. Although such a procedure would lose the reionization bump, robustness to foreground threshold variation of any $r$-detection could be well-demonstrated. Apart from its many other virtues, Planck should be very good for this. 

%---------------------------------------------------------------
\subsection{Relation to Spider}\label{sec:cfSpider}

The same strategy of using many fields with the lowest foregrounds to make up the total $ f_{\rm{sky}}$ may also prove useful for Spider-like 
experiments (such as the ground-based ABS). We showed that splitting $ f_{\rm{sky}}$ into four patches with fixed integration time and the instrument noise results in only a small loss in $r$-sensitivity because $\sigma_r ( f_{\rm{sky}} )$ has a relatively wide single-patch  minimum. 
How many polarization-foreground-clean patches there are is still to be determined. 

Although the specifications we chose for "Spider-like"   was motivated
by a bolometer array experiment feasible with current technology, our
forecasts should not be taken as realistic mocks of the true Spider
which is under development, and for which a number of campaigns are
envisaged (see the footnote under Spider-like in Table~\ref{experiments}). 
The techniques used here have, however, already been applied in Spider forecast papers using more realistic statistically inhomogeneous noise, scanning strategies and observational durations, e.g.,  in \cite{fli10} and \cite{fra11}.   On an $f_{\rm{sky}} \sim 0.1$, $r_{\rm fid}=0.01$ simulations, we compared \cite{fra11} non-uniform noise modulated spatially by the scanning strategy's number-of-hits-per-pixel with uniform white noise with the same integrated noise power. Although the deviation in the standard deviation of the noise {\it rms}  was about a factor of two times the mean noise {\it rms}, with largest impact near the scanning boundaries, we found very similar results for the posterior, showing this paper's  conclusions are insensitive to our use of uniform white noise.  (Of course the foreground noise radically alters the whiteness, and this of course has been included by us, but only in a statistically isotropic way --- the Galactic latitude dependence  breaks this isotropy just as the pixel hits do.)  
%------------------------------------
\begin{figure}[h]
\begin{center}
\includegraphics[scale=0.55]{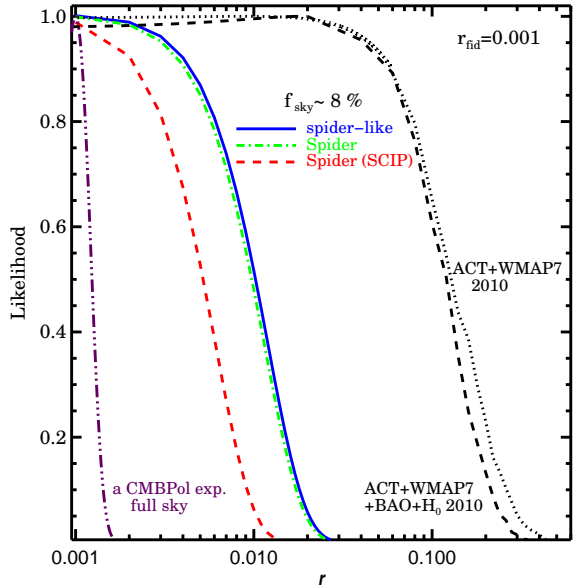}
\end{center}
\caption{The $r$-likelihood curve for the Spider-like experiment
  (which is the default experiment used in this paper) with $r_{\rm fid}=0.001$ and $f_{\rm{sky}}=0.08$
  is contrasted with proposed stages in balloon-borne experimenting
  with an actual  Spider focal plane. The one labeled as Spider
  corresponds to the 
  actual, more recent Spider proposal  with two flights described in
  \cite{fra11} (see the footnotes of table~\ref{experiments}). The SCIP envisages
  three subsequent flights of the Spider payload. We see that the future sensitivity may
  exceed this paper's forecasted constraints. These Spider likelihood
  curves have been contrasted with the current limit on $r$ from CMB (ACT+WMAP7)
  alone and from CMB with measurements of $H_0$ and BAO \citep{dun10}.
  The marginalized $1D$ likelihood curves are based on the publicly
  available chains
  {http://lambda.gsfc.nasa.gov/product/act/act\_chainsv2\_get.cfm}
  binned into 50 bins, and Gaussian-fitted to plot the very small $r$
  region where not enough points were available.  
  These current and near future constraints are compared to the expectation form the next generation of 
  space CMB mission. As an example, we used the results of simulations
  for a full sky CMBPol experiment (see table~\ref{experiments}) again with $r_{\rm
    fid}=0.001$, which gives $2\sigma_r \sim 0.0004$, comparable to the
 forecasted errors from PIXIE $2\sigma_r \sim 0.0004$ and COrE $2\sigma_r \sim 0.0007$.  }
\label{fig:spider_apra_scip}
\end{figure}
%----------------------------------
In \S~\ref{sec:rtau}, we showed that in the absence of foregrounds our Spider-like case 
could achieve $2\sigma_r \approx
0.02$ over a broad range of $f_{\rm{sky}}$.  The
$r$-posteriors shown in Figure~\ref{fig:spider_apra_scip} were made
with the numerical codes described here, 
for the Spider experiment as envisaged in \cite{fra11} (labeled as
``Spider'' in the plot), and for an even more ambitious campaign of
subsequent flights of the Spider instrument, as proposed for SCIP. We
see that the performance of the experiment with Spider-like
specifications used in this paper is very close to the actual Spider. 
 A different foreground model
used in \cite{fra11} for $f_{\rm{sky}}\sim 0.1$  led to a similar  $\sim 50\%$ error degradation. 

%------------------------------------------------------------------
\subsection{History and Forecasts of $r$ Constraints}\label{sec:sigrhistory}

When the large angle CMB anisotropies were first detected with COBE
DMR, the broad-band TT power amplitude ($\ell \lesssim 20$), with
wavenumbers $k^{-1} \gta 1000$ Mpc, was related to the linear density
power spectrum amplitude at the radically different $k^{-1} \sim 6$
Mpc scale, assuming a nearly scale-invariant primoridial spectrum:
$\sigma_8 \approx  0.85e^{-(\tau -0.1)}/ \sqrt{1+0.6r} \times
1^{0.7}_{-0.6}$ for typical $\Lambda$CDM parameters popular in mid nineties, $\Omega_\Lambda \sim 2/3, h \sim 0.7$ \citep{bon96}, rather similar to the values now. Requiring $\sigma_8 > 0.7$ to get reasonable cluster abundances at zero redshift -- a venerable cosmological requirement from the 80s -- gives a rough constraint on $r$ from the COBE data in conjunction with large scale structure (LSS) data:  $2\sigma_r < 1$ for current $\tau$ values -- but $\tau$ only had an upper limit until WMAP1, with a more accurate determination waiting until WMAP3. 

The first 2003 WMAP1 constraint on $r$ from  $TT$ and $TE$ CMB-only data (with weak priors) was $2\sigma_r < 0.81$, reducing to $2\sigma_r < 0.64$ with the WMAP3  $TT, TE$ and $EE$ data, and other $TT$ CMB data available in 2005. It  decreased to 0.31 with the LSS data of the time \citep{mac06}.  The most recent $r$-constraint from the low $\ell$ amplitude and shape of the $TT$ and $EE$ spectra from WMAP7+ACT is the upper limit $2\sigma_r \sim 0.25$, reducing to $0.19$ when LSS is added \citep{dun10}. 

To make a further leap awaits an effective $BB$ mode constraint. As we
have seen, Planck can give 0.015-0.05, Spider 0.014-0.02. The  COrE
satellite proposal \citep{core11} suggests better than a 3-sigma
detection could be made for  $r_{\rm fid}$ above 0.001 with bolometer
arrays in space. The PIXIE satellite proposal \citep{kog11} claims
$2\sigma_r \approx 4\times 10^{-4} $ is achievable with Fourier
Transform Spectrometry.   Applying our methods to CMBpol
specifications \citep{bau09} we get $2\sigma_r \approx 4\times 10^{-4} $ for $r_{\rm
  fid}=0.001$ and $2\sigma_r \approx 1.2\times 10^{-4} $ for $r_{\rm
  fid}=0.0001$.  If $r_{\rm fid}$ is as large as 0.12, as in the
simple $m^2\phi^2$ chaotic inflation, we get $2\sigma_r \approx  0.015$
(and $2\sigma_{n_{\rm t}}\approx 0.07$ encompassing the consistency input  of
$n_{\rm t}=-0.015$). For a noiseless all-sky experiment, hence with errors
from cosmic variance only, we get $2\sigma_r \approx 10^{-4}$ for
$N_{\rm side}=128$ for tiny $r_{\rm fid}$. It is unclear at this time how much inexact foreground subtraction and lensing noise will limit $r$ determinations in these ideal cases. 

%-------------------------------------------
\subsection{The 1D Shannon Entropy of $r$}\label{sec:S1fr}

We have described another way to cast the improvements expected in
$r$-estimation as experiments attain higher and higher sensitivity,
the marginalized 1D Shannon entropy $\Delta S_{1{\rm f}}(r)$ for $r$. This
measures the (phase-space) volume of $r$-space that the measurement
allows. It is obtained by direct integration over the normalized 1D
likelihood for $r$, with all non-Gaussian features in the likelihood
properly included. We have found in practice that $\Delta S_{1{\rm f}}(r)
\approx \Delta \ln [\sigma_r \sqrt{2\pi} ]$, with $\sigma_r$
determined by the forced Gaussianization described in the paper, works
quite well, so in a way we are just restating the error improvements
in the information theoretic language of bits.    

We use the WMAP7+ACT $TT, TE$ and $EE$ $2\sigma_r \sim 0.25$ \citep{dun10}
constraint for our baseline. The first WMAP1 constraint in 2003 \citep{spe03}, with
$\Delta S_{1{\rm f}}(r) =1.70$ bits had, of course, higher information
entropy. Here, as in the abstract, we have translated from nats to
bits.
The recent WMAP7+SPT results \citep{kei11} with $2\sigma_r \sim 0.21$ give a
slight decrease in the entropy ($\Delta S_{1{\rm f}}(r) =-0.25$) compared to the baseline.  
The asymptotic perfect noiseless all-sky experiment gives (the
somewhat $r$-dependent) $\Delta S_{1{\rm f}}(r) \approx -11 $ bits, the
limit on obtainable knowledge from the CMB.  The proposed post-Planck
COrE, PIXIE and  CMBPol-like experiments claim up  to -9 bits. For the Spider-like experiments forecasted here, the foreground-free decrease is -4.2  bits (and  -3.6 bits with a 95\% effective component separation). Thus balloon-borne and ground-based experiments with large arrays making deep 
surveys focussing on a relatively clean few-percent of the sky yield tensor information at least comparable
to shallow and wide surveys and are a powerful step towards a near-perfect  deep and wide satellite future. 

We would like to thank our many Spider, ABS and Planck collaborators for many stimulating discussions about the experimental assault on CMB tensor mode detection. We would like to thank William C. Jones  for his helpful comments on the text. We thank Marc Antoine Miville Desch\^enes for advice and aid on foregrounds. 
Support from NSERC, the Canadian Institute
for Advanced Research, and the Canadian Space Agency (for PlanckÐHFI and Spider work)
 is gratefully acknowledged. Part of the research described in this paper was carried out at the Jet Propulsion Laboratory, California Institute of Technology,  under a contract with the National Aeronautics and Space Administration. The large matrix computations were performed using the SciNET facility at the University of Toronto. Some of the results in this paper have been derived using the HEALPix package \citep{gor05}, {http://healpix.jpl.nasa.gov}. 

%*****************
\bibliography{r_measurement}
\bibliographystyle{apj}

\end{document}